\documentclass{raa}           
\usepackage{graphicx,times}
\usepackage{natbib}
\usepackage{amssymb,amsmath}
\usepackage[dvips]{epsfig}
\usepackage{pstricks}
\usepackage{pst-plot}
\usepackage{pstricks-add}
\bibpunct{(}{)}{;}{a}{}{,}

\usepackage[letterpaper=true,pagebackref=true]{hyperref}
\hypersetup{pdftitle = The title of my PDF, pdfauthor = My name, pdfsubject= The subject, pdfkeywords = keyword1 keyword2 keyword3} 
\hypersetup{colorlinks = true, linkcolor = green, anchorcolor = red, citecolor = blue, filecolor = red, pagecolor = red, urlcolor = red}
\begin{document}

\title{Using Cartesian slice plots of a cosmological simulation as input of a convolutional neural network$^*$
\footnotetext{$*$ Departamento de Investigaci\'on en F\'{\i}sica, Universidad de Sonora.} }

\volnopage{ {\bf 20XX} Vol.\ {\bf X} No. {\bf XX}, 000--000}

\setcounter{page}{1}

\author{Guillermo Arreaga-Garc\'{\i}a\inst{1}}
\institute{Departamento de Investigaci\'on en F\'{\i}sica, Universidad de Sonora. \\
Apdo. Postal 14740, C.P. 83000, Hermosillo, Sonora, Mexico. \\
guillermo.arreaga@unison.mx\\
{\small Received 20XX Month Day; accepted 20XX Month Day}
}
\abstract{Using a uniform partitioning of cubic cells, we cover the total volume of 
a $\Lambda$CDM cosmological simulation based on particles. We define a visualisation cell as a spatial extension 
of the cubic cell, so that we collect all simulation particles contained in this visualisation cell 
to create a series of Cartesian plots in which the over-density of matter is clearly visible. We then 
use these plots as input to a convolutional neural network (CNN) based on the Keras library and TensorFlow for 
image classification. To assign a class to each plot, we approximate the Hessian of the gravitational 
potential in the centre of the cubic cells. Each selected cubic cell is then assigned a 
label of 1,2 or 3, depending on the number of positive eigenvalues obtained for the Householder reduction 
of the Hessian matrix. We apply the CNN to several models, including two models with different visualisation 
volumes, one with a cell size of type L (large) and the other with a cell type S (small). A third 
model that combines the plots of the previous L and S cell types. So far, we have mainly considered 
a slice parallel to the XY plane to make the plots. The last model is considered based on 
visualisations of cells that also include slices parallel to the ZX and ZY planes. We find that the accuracy in classificating 
plots is acceptable, and the ability of the models to predict the class works well. These results 
allow us to demonstrate the aim of this paper, namely that the usual Cartesian plots contain enough information 
to identify the observed structures of the cosmic web.}

\authorrunning{Arreaga-Garc\'{\i}a}            
\titlerunning{Cartesian plots to classify matter with CNN}  
\maketitle
\section{Introduction}
\label{sec:int}

With the advent of artificial intelligence and machine learning, a convolutional neural network (CNN) can 
recognise complex patterns in plots. For example, the simple MNIST-CNN described by \citet{chollet} recognizes a handwritten 
number from a dataset of thousands of images. Among the many CNNs available, see \citet{alexnet}, \citet{GoogleNet} and \citet{resnet}, we 
consider in particular the CNNs described by \citet{bagnato}: one CNN is trained to recognize 9 classes of sports 
from a dataset of thousands of sports images and the other CNN is trained to classify images of dogs and cats from a dataset of 
thousands of dog and cat images. 

In this work, we train the image classification CNNs described by \citet{bagnato} with a dataset of 
rendered 2D plots commonly presented in particle-based numerical simulations to show the simulation 
results. These plots are mainly two-dimensional Cartesian coloured iso-density plots for a slice of particles 
from the simulation volume. Particularly in the early simulations, the comparison of the results of particle-based simulations 
with observations was usually done using plots of a Cartesian slice through the distribution of simulation 
particles, usually in two dimensions (plots in three dimensions were also produced, but less frequently as they 
became increasingly difficult to produce). It should be noted that an essentially subjective visual 
inspection of these 2D plots was used to perform morphology comparisons of matter structure between 
different simulation models; see for example, \citet{weinberg}. 

In supervised machine learning, the dataset of plots must be supplemented by a dataset of 
classes, i.e. each plot must have a class, which is simple a label. One could choose an attribute of interest 
to determine the class of a plot. In the case of the CNN described by \citet{bagnato}, there are 9 classes for the first CNN and 
only two classes for the second CNN, namely dog or cat, so each image must be labelled as dog or cat. After 
the training process, these CNNs can successfully distinguish the type of sport or whether the given image corresponds 
to a cat or a dog. 
  
To give a physical content to the dataset of the plots of this paper, we consider
\citet{hahn}, who performed N-body particle-based cosmological simulations and considered only 
dark matter haloes with at least 300 simulation particles in the snapshot at a redshift $z=0$. \citet{hahn} counts the 
number of positive eigenvalues of the Hessian matrix, which is defined as 
$T_{ij}=\frac{\partial^2 \Phi}{\partial x_i \, \partial x_j}$, where $\Phi$ is the gravitational potential. The number 
of positive eigenvalues can be 0, 1, 2 or 3 and corresponds to a void, a sheet, a filament and a cluster, which 
represent the most common distribution of matter in the cosmological simulation.

To follow the method described by \citet{hahn}, we define a uniform partition of cubic cells covering the 
total volume of a particle-based $\Lambda$CDM cosmological simulation, and determine the eigenvalues of the Hessian of the gravitational 
potential. We then count the number of positive eigenvalues in each cubic cell and use this number as the label for the plots 
constructed for that cell.

If one can successfully train the CNN with this kind of dataset, what would be the prediction that the CNN 
could make ?. In this case, it would be possible to recognize and distinguish between filaments, walls and clusters. The aim of this paper is to show that the usual Cartesian plots contain enough information 
to identify the observed structures of the cosmic web. To be sure that this is the case, we need to make calculations. 
Let us firt put the work of \citet{hahn} in context to illustre the importance of this problem.

The investigation of the spatial distribution of matter in the universe on scales of a few Mpc dates back to the 
1930s; see \citet{shapley}. Later, using galaxy catalogues, it was 
discovered that on these scales there were filaments, voids and sheets in which the galaxies 
were housed; see \citet{geller}. With the advent of numerical simulations, it became clear that the 
large-scale structure of the universe consists of a network of dark matter that is interconnected in 
complex ways and was aptly named the cosmic web; see \citet{bond}.

The quantitative analysis of the large scale distribution of matter has a long and interesting history. For 
example, \citet{shandarin} tried to find patterns in the distribution of galaxies, and defined a 
function $B(r)$ to represent the average number of galaxies within a sphere of radius $r$. They obtained a percolation 
radius that depended on the distribution of galaxies in the sample cube. From this result, they were able to deduced that in the universe 
the galaxies are distributed according to a network structure. Shortly afterwards, \citet{barrow} proposed a graphical algorithm, the 
so-called minimal spanning tree to determine intrinsic patterns in point data sets. This algorithm was applied to two and three 
dimensions datasets to compare the distributions of real galaxy catalogues with random distributions.

\citet{gott} constructed density maps to investigate the relationships between the high and low density regions of a smoothed 
distribution of galaxies from a galaxy catalogue. Finally, they presented a model that assumes a sponge-like structure of 
the Universe. At that time, the distribution of galaxies was also modelled as a homogeneous 
fractal, and several attempts were made to calculate its Hausdorff dimension, see \citet{martinez}. \citet{babul} introduced the 
so-called structure functions to measure certain geometric properties, such as sphericity, prolatency and oblatency. With the help 
of these structure functions, they began to quantify the morphology of the matter distribution through simulations.

We should also mention some recent papers devoted to the study of the cosmic web. \citet{libeskind} presents a very 
comprehensive overview of the various methods that have been developed to classify and identify the different matter elements of the 
cosmic web, such as: (i) graph and percolation techniques (see \citet{shandarin} and \citet{alpaslan}) using a method known as 
the adapted minimal spanning tree (MST); (ii) stochastic methods (see \citet{leclercq}, \citet{tempel2014} 
and \citet{tempel2016}) based on the Bisous method; (iii) geometric, Hessian-based methods (see \citet{aragon2007} and 
\citet{hahn}; \citet{aragon2007} presented a multi-scale morphology filter to recognize three different structural configurations of the 
cosmic web, namely blobs, walls and filaments); (iv) scale-space multiscale Hessian-based methods (see \citet{cautun} with 
the NEXUS+ method); (v) topological methods (see \citet{colombi} and \citet{aragon2010} with the Spineweb method and \citet{sousbi} with 
the DisPerSE method); and (vi) phase-space methods (see \citet{shandarin2011} and \citet{rama} with the multi-stream web 
analysis (MSWA) method, and  \citet{falck2012} and \citet{falck2015} with the ORIGAMI method). 

The $\beta$-skeleton method 
has been used successfully in various areas of pattern recognition. In 
particular, this method can be used to reconstruct images from partial data or from data concerning only  
the contours. It is based on an algorithm that makes it possible to associate a graphical structure 
with a set of points in a discrete data set. The data can be in three or two spatial dimensions. This graphical 
structure determines the connectivity of the points in the dataset and depends on the value of a 
parameter called $\beta$. \citet{fang} has applied this method 
to the large-scale structure of the Universe and has succeeded in recognizing the filaments 
of the cosmic web; see also \citet{garcia}.

\citet{hoffman} and \citet{forero} have introduced a 
threshold eigenvalue, so that the counting of positive eigenvalues proposed by \citet{hahn} is extended to 
counting the number of eigenvalues above a certain threshold eigenvalue to determine the class of the given halo.
\citet{aragon2019} presented a cosmic web classification using convolutional neural networks (CNNs) with a U-net architecture, 
previously used in the classification of 2D medical images and 3D data cubes of medical interest. The 
density function is smoothed from the discrete distribution of the simulation particles to obtain a continuous density 
field based on the second-order local variations of the density field encoded in the Hessian matrix. A very 
important step is the determination of the threshold eigenvalue. \citet{aragon2019} has proposed an automated algorithm, but 
the best results are obtained by visual inspection of the algorithm results.

Recently, \citet{inoue} utilised a CNN fed with plots of the distribution of galaxies 
and particles from the IllustrisTNG simulations. They used the velocity gradient tensor to 
obtain the eigenvalues above a threshold to obtain the classification scheme. They also considered
a set of four labels:  voids, sheets (also call walls) , filaments and knots (also called clusters). Their models 
can classify simulated galaxies with an accuracy (macro-averaged f1-score) of 64 percent. To obtain the 
cosmic structure classification, they used a coverage of 256$^3$ cells and considered only dark matter particles.

Considering the papers of \citet{aragon2019} and \citet{inoue}, the calculations proposed in this paper are not new.
And after mentioning some of the many existing methods for classifying the elements of the cosmic web, we emphasize that our method 
does not provide any element that improves these methods. The novelty of this work is that we show that the plots commonly used 
to see the results of a cosmological simulation can also be used to classify the elements of the cosmic web network using a CNN.  

The structure of this paper is as follows. In Section \ref{subs:simulation} we describe the cosmological model. In Subsection
\ref{subs:code}and we provide some details about the simulation. 
In Subsection \ref{subs:cells} we describe the coverage of the cubic cells and the method used to 
approximate the Hessian in each selected cubic cell.  In the Appendix \ref{app:compacells} we complement this Subsection \ref{subs:cells}. In Subsection \ref{sec:consistencytest} we present a consistency test 
of the calculation described above. In Subsection \ref{sec:resultsdistri} we present a characterization of the matter 
content of the cells. In Subsection \ref{subs:training} we 
describe the development of the sets of training and validation plots for two models with different 
visualisation volumes. In Subsection \ref{subs:network} we describe the CNN and in the Appendix \ref{app:network} we present 
an alternative CNN. In Section \ref{sec:results} we show the results, which include some reports on the training and validation analysis 
(in Subsection \ref{sec:report}) and a reports on the prediction ability using a confusion matrix analysis ( in 
Subsection \ref{sec:reportpredi}) for each CNN model. Finally, in Sections \ref{sec:dis} 
and \ref{sec:conclu}, we discuss the relevance of our results with respect to the results of previous papers and 
make some concluding remarks. 
\section{The physical system and computational considerations}
\label{sec:phy-sys}
\subsection{The simulation}
\label{subs:simulation}

We consider a cubic box with a side length given by $L=75$\ Mpc. The initial content
of matter is characterised by $\Omega_m=$ 0.2726 and the content of dark energy is
$\Omega_{\Lambda}=$0.7274, so that $\Omega_m+ \Omega_{\Lambda}=$ 1.0, corresponding to a flat model of the universe, expanding 
with a Hubble parameter H$_0=100 \, h$ km s$^{-1}$ Mpc$^{-1}$.  $h$ is given by $h=0.704$. The baryon mass is characterised 
by $\Omega_b=$ 0.0456. These values have 
been chosen from \citet{planck} and \citet{planck2016}. 

The density perturbations of this cosmological model have been generated with 
the publicly available code \citet{n-genic}, so that an initial power 
spectrum $P(k)$ can be constructed by moving the simulation particles according
to the linear spectrum defined by \citet{eisenstein}. The power spectrum normalisation was fixed at a 
value of $\sigma_8$=0.9.

It should be noted that for each dark-matter particle there is a
gas particle. Thus, the number of dark-matter (DM) particles $N_{\rm DM}$ and the 
number of gas (G) particles $N_{\rm G}$ are both equal to 11,239,424.
Therefore, the particles have masses given by $m_{DM}= 2.3 \times 10^{9} M_{\odot}$ and
$m_{G}= 4.7 \times 10^{8} M_{\odot}$, respectively. The average mass density is 
given by $\rho_0=$ 5.12 $\times$ 10$^{-30}$ g cm$^{-3}$ and the
initial and final redshifts are fixed at $z=127$ and $z=0$, respectively.

\subsection{The evolution code}
\label{subs:code}

The simulations used in this paper are evolved with the particle-based code 
Gadget2; see~\citet{springel}. Gadget2 implements the Monaghan-Balsara form
of the artificial viscosity \citet{balsara1995}, so that the strength of the
viscosity is regulated by setting the parameter $\alpha_{\nu} = 0.75$ and
the parameter $\beta_{\nu}=\frac{1}{2} \times \alpha_v$. The Courant factor has been fixed at $0.1$.

The time evolution of the simulation up to $z=0$ required a little more than 80 hours; it ran on 60 
processors in the cluster Intel Xeon E5-2680 v3 at 2.5 Ghz at Laboratorio Nacional de Supercomputo (LNS) of 
the Benemerita Universidad Autonoma de Puebla (BUAP). The computational method described in 
subsection \ref{subs:cells}, the results of which are described in subsection \ref{sec:results}, was 
applied only to the last output, that is, the snapshot at a redshift $z=0$.
\subsection{The cubic cells and the Hessian}
\label{subs:cells}

With the aim of investigating the grouping properties of proto-galaxy clusters, \citet{arreaga2021} used partitions of two 
sizes, with 64$^3$ and 128$^3$ cubic elements, to search for matter structures that had already gravitationally collapsed. We chose 
the first partition, so that the total 
number of identical cubic cells into which the entire simulation volume is divided is therefore 64$^3 \, \equiv $ 262144. It should be 
noted that the side length of each cubic cell is $2*\Delta_H=1.17$ Mpc 
where $\Delta_H=0.585$ Mpc.

A very small cell size would allow us to detect only the core of a protogalaxy, and almost all cores would be isolated, while a very large 
cell size would allow us to detect many structures within the cell, almost all of which would be connected, but it would be difficult to 
label the cell. Remember that we are trying to label each cubic cell according to the type of matter it contains. With this 
objective in mind, the size chosen for the cubic cell can be physically justified as follows.

In the percolation technique, this idea of the connectivity of protogalaxies can be quantified by the percolation 
radius, \citet{zeldovich2}. In general, the connectivity of galaxies decreases with increasing percolation radius, a similar 
behaviour to the two-point correlation function, where the test radius $r_0$ is about 5 Mpc.

Let us consider the distribution of galaxies around the Virgo cluster. In the percolation technique, a suitable radius for modelling the 
Virgo cluster using a cylinder would be 2.2 Mpc, for Perseus 2.9 Mpc, and for Coma 2.5 Mpc. Therefore, a radius 
of 2 Mpc can therefore be assumed to cover this type of distribution of galaxies with a cylinder model. This model allows us to 
consider a chain of galaxies that are part of a single matter structure. In this case, we can in principle assign a label to 
a grid cell containing this type of connected matter structure.
 
To this physical justification for the size of the cubic cell we can also add a technical justification. The resolution 
of the simulation described in Section \ref{subs:simulation} is sufficient to see the matter structures considered 
above, i.e. protogalaxy clusters. Furthermore, the plots generated for each selected cubic cell must have enough 
features to be useful as input for the neural network. For this reason, the cell size must not be too small for matter 
structures to be recognised, but at the same time, the cell size must not be too large to obtain a label. Therefore, a test cell size 
in the range 1-7 Mpc can be used, and plots are then created to assess whether they are useful. Due to this 
indeterminacy of the cell size, we use two cell sizes in this study. 

We continue with the description of the following steps. We determine the number of 
simulation particles (including dark matter and gas particles) that 
are in each cubic cell. If the actual cell has more particles than a certain threshold 
number $N_t$, then we approximate the Hessian for that cell as follows. We need to calculate 
$T_{ij}=\frac{\partial^2 \Phi}{\partial x \, \partial y}$, where
$\Phi$ is the gravitational potential. We use the particle approximation $\Phi(\vec{r})=\Sigma 
\Phi(\vec{r}_i) \, W(\vec{r}-\vec{r_i} )$, where $W$ is a window function. The second 
derivatives are approximated by 
$f''(x_0)= \frac{ f(x_0+h) + f(x_0-h)-2f(x_0)}{h^2} 
+\frac{h^2\, f^4(\zeta)}{12}$ for $\zeta \, \in \, (x_0-h,x_0+h)$, where $h$ is a small
increment, here assumed as $h=\Delta_H/100$. 

We then use our own code to calculate all components of the symmetric tensor $T_{ij}$ in this 
particle approximation. The subroutine tred2.c is used to reduce the matrix $T_{ij}$ to a 
diagonal form, and the subroutine tqli.c is applied to find the eigenvectors and eigenvalues of the 
matrix $T_{ij}$. Both subroutines were taken from \citet{recipes}. 

In this approach, the particles located in the neighbouring cells are not taken into account when calculating the Hessian for a 
given cell. Therefore, we do not quantify the effects of including neighbouring cells in the Hessian diagonalisation. This calculation 
was performed in this way to simplify the process. However, we have performed a comparison of the results obtained when 
calculating the Hessian with two differently sized cells as can be seen in the Appendix \ref{app:compacells}.

This code runs on all cubic cells, so we get the centre of the 
selected cells, the number of particles contained in each selected cell and the number of positive eigenvalues of the 
Hessian. We mentioned in Section \ref{sec:int} that we only considered the cells whose number of 
simulation particles is above a certain threshold number. We need to note the importance of the threshold number 
of particles as a free parameter. For example, using $N_t=15$ the total number of cells was 
77367, with the following distribution: there were 59544 cells with three positive eigenvalues; there were 
7462 cells with two positive eigenvalues and there were 10361 cells with only one positive eigenvalue. 

If you increase this parameter to $N_t=150$ particles, these numbers change to 14515, 145 and 133 
cells, for three, two and one positive eigenvalue, respectively. In both cases, the number 
of class 3 cells is much larger than the number of class 1 and 2 cells. In this paper, we have set the
threshold for the number of particles to 150.  

By setting a minimum number of particles per cell, we can ensure that the plot of each cell has sufficient visual content 
for the analysis to be meaningful. This minimum limit for the number of particles is equivalent to setting a minimum limit for the 
density of the cells. For example, when a plot of a simulation box is created, only the particles whose density is greater than a certain 
limit are selected to facilitate the visualisation of the dense matter of a simulation.

Finally, we order these datasets in terms of the number of simulation particles 
contained in the cell, from the highest to the lowest value to generate training plots with the best possible resolution, that 
is, with more simulation particles, as we will explain in Section \ref{subs:training}. The cells with 
only a few particles generate plots with low resolution and were therefore discarded.

\subsubsection{Consistency tests}
\label{sec:consistencytest}     

As a consistency test for this code, the halo finder RockStar described by \citet{rockstar} was used to locate the 
haloes of matter determined by constructing maximum isodensity curves. We placed cells 
in the halo centres located by Rockstar and then executed the code described in this Section \ref{subs:cells}. 
We found that in this test, the output of the code indicates 
that all cells are of class 3, i.e. they have three positive eigenvalues. So, if you use the over-density 
centres obtained by the halo finder code Rockstar, our code will not find any class 1 or 2 cells, as 
expected. In Section \ref{sec:dis}, we will compare our results with those obtained with the code 
described by \citet{forero}. 

\begin{figure}
\begin{center}
\begin{tabular}{c}
\includegraphics[width=3 in]{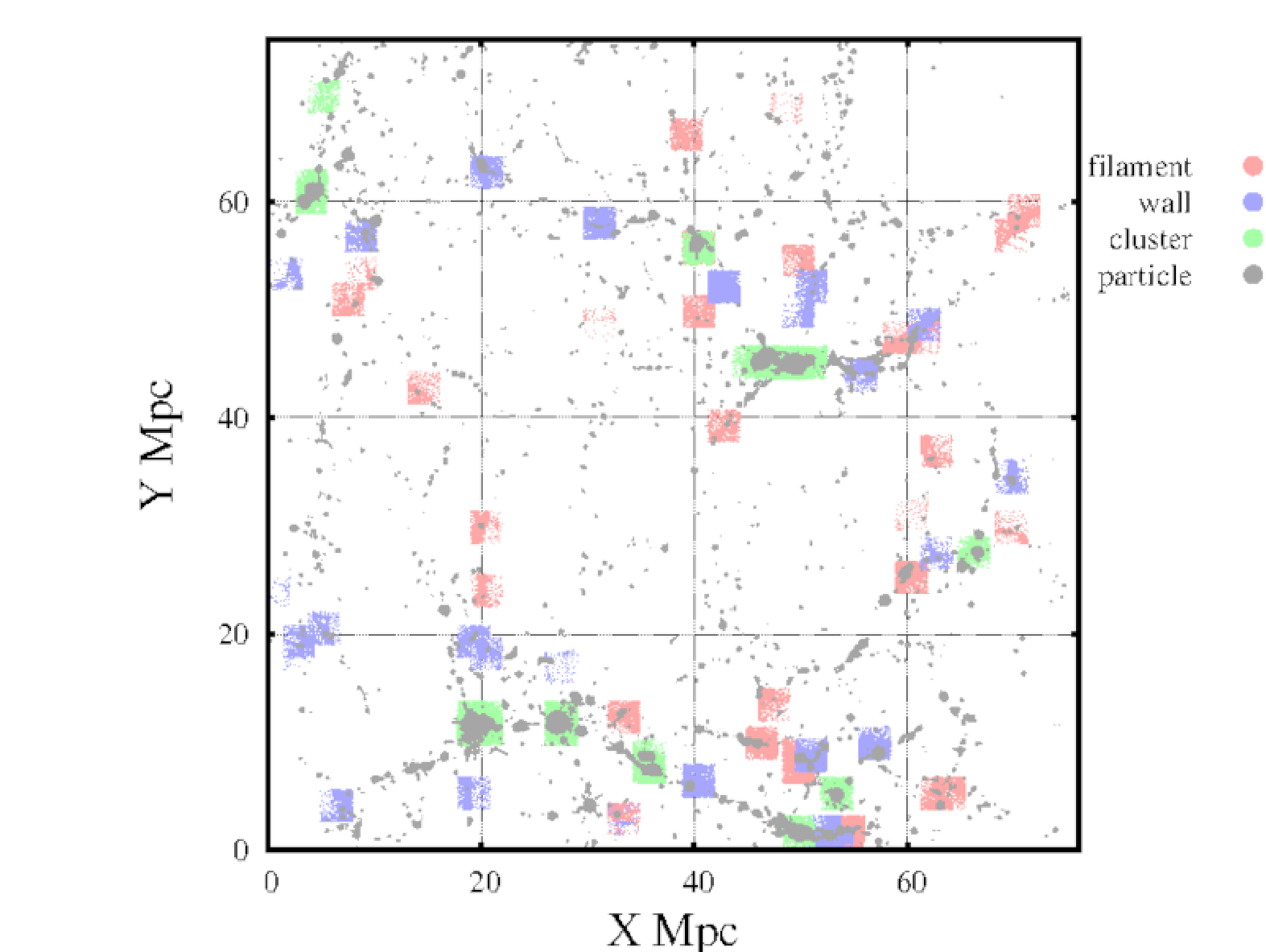}\\
\end{tabular}
\caption{\label{DistriEstruc} This Figure shows a comparison between the spatial distribution of the three classes of matter structure 
and the simulation particles located in a slice. The plot is a 2D projection of a 3D slice 
of the simulation box, with a width of 9 Mpc in the coordinate perpendicular to the plane indicated on the axes of each panel.
The height of the middle plane is located at  37.5 Mpc in the perpendicular coordinate.}
\end{center}
\end{figure}

In addition, in Fig. \ref{DistriEstruc} we show the distribution of some simulation particles extracted from a section of the simulation box 
as well as the cells with matter structures classified as filaments, walls or clusters. It should be emphasised that this figure only shows 
simulation particles. All particles that are located in a grid cell and have been assigned a class label are collected in a file, and 
coloured according to their class, as shown at the top right of each plot in the Fig. \ref{DistriEstruc}. 

These particles are superimposed on the simulation particles so that we can search for matches. It can be seen that there is a good match for 
clusters (class 3). As expected, the identification of filaments and walls by the human eye is not easy. Therefore, we are investigating 
whether a CNN can perform this classification task across the entire simulation volume.

\subsubsection{The distribution of cells in terms of the mass}
\label{sec:resultsdistri}   

\begin{figure}
\begin{center}
\begin{tabular}{cc}
\includegraphics[width=2.7 in]{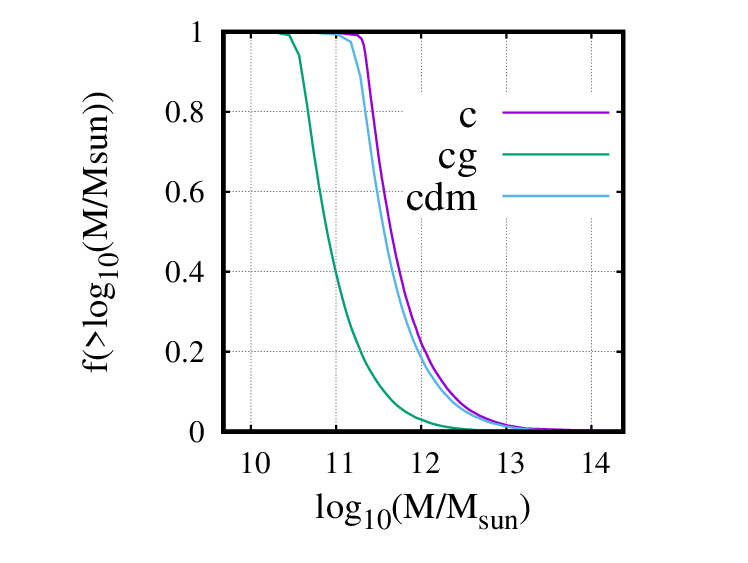} & \includegraphics[width=2.7 in]{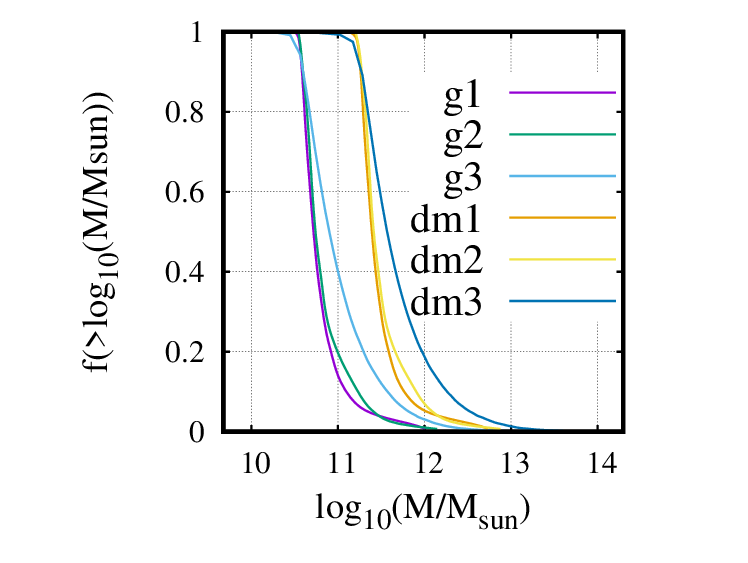} \\
\end{tabular}
\caption{\label{plotsFM} Distribution of cubic cells versus the mass that they contain. We show on 
the vertical axis the fraction of cells with a mass greater than (or equal) to the given mass on 
the horizontal axis. We take into account both components, the gas and dark matter, irrespective of 
the number of eigenvalues of each cell (left panel).  We also consider separately the gas and 
dark matter in the cells according to the number of eigenvalues of each cell (right panel). The classes here are 
referred to as 1,2 and 3. The curves in the right panel are labelled as g1 (dm1), which refers to 
gas (dark matter) in the cells of class 1 (with only one positive eigenvalue); and g2 (dm2) and g3 (dm3), which
analogously refer to gas and dark matter in cells with two and three positive eigenvalues, respectively.}
\end{center}
\end{figure}

To conclude this Section \ref{subs:cells}, we present the distribution of cubic cells in terms of the mass 
they contain. As we mentioned above, we have consider the gas and dark matter components 
when calculating the Hessian. Now we consider these components separately to investigate their distribution in 
the structure classes. First, in the left panel of Fig. \ref{plotsFM} we calculate the fraction 
of cells with a mass greater than or equal to the given mass indicated on the horizontal axis of the panel, regardless of the class 
of the cell. There are three curves in this panel: one is labelled 'c', which means that all the matter contained in 
the cubic cell has been taken into account. For the curve labelled 'cg', only the gas contained in the cell has been
considered and analogously, for the curve labelled 'cdm', only the dark-matter contained 
in the cell has been taken into account. As expected, we see that the dark-matter dominates the matter 
content of the cells.

In the right panel of Fig. \ref{plotsFM} we show the distribution of the cells and the classes of the 
cells. The curves are labelled g1 (dm1), which refer to gas (dark matter) in the cells of class 
1 (with only one positive eigenvalue), and g2 (dm2) and g3 (dm3), which analogously refer to gas and 
dark matter in cells with two and three positive eigenvalues, respectively. In this panel we see that 
there are two groups of curves, one group for the gas component and the other group for the dark-matter component.

It should be noted that there is a threshold for the number of particles that must be reached for a cell to be 
selected; see Subsection\ref{subs:cells}. For this reason, most curves in each group coincide 
at small masses and are separated at large masses. It can be seen that the structures in class 3 (the clusters) have the largest 
masses in both groups, 
i.e. for gas and dark-matter components, as reported by \citet{veena}. However, in the upper-left corner 
of each group of curves, it can also be seen that the smallest masses are also detected 
for class 3 cells. The curves for class 2 and class 3 cells agree in each group for small masses, but 
are slightly separated for large masses.  
  
\subsection{Training plots and the CNN models}
\label{subs:training}

In this section we consider the visualisation cells as follows. We take the coordinates of each cell centre $x_c,y_cz_c$ and determine all simulation 
particles whose coordinates are within the interval $x \, \in \, [x_c-\Delta_L , x_c+\Delta_L]$ where $\Delta_L$ is a 
width, that we will define below and analogous relations for the $y$ and $z$ coordinates are not shown. All these simulation 
particles are considered and plots are created for this visualisation 
volume, which has a side length of $2* \Delta_{L_L}=3.0$, where $\Delta_{L_L}= 1.5$ Mpc. This visualisation cell is called an L cell.  

In the Subsection \ref{subs:cells} we mentioned that to approximate the Hessian, we considered all particles 
in the cubic cell H with a side length of $2* \Delta_H=1.17$ Mpc, so that the visualisation volume is 
larger than the volume of the Hessian. This must be so because we expect to capture extended objects such as filaments 
and walls, which can be better sampled by a wider grid. 

We also consider a smaller visualisation volume where the side length is $2* \Delta_{L_S}=1.5$ where $\Delta_{L_S}= 0.75$ Mpc. This 
visualisation cell is called S cell. To distinguish these models, we use the labels 'L' and 'S' respectively, as shown 
in Table \ref{tab:numeroimagenes}. In Section \ref{sec:results} we present the results of these runs with different sizes 
for the visualisation volume.

\begin{figure}
\begin{center}
\begin{tabular}{cc}
\includegraphics[width=2.7 in]{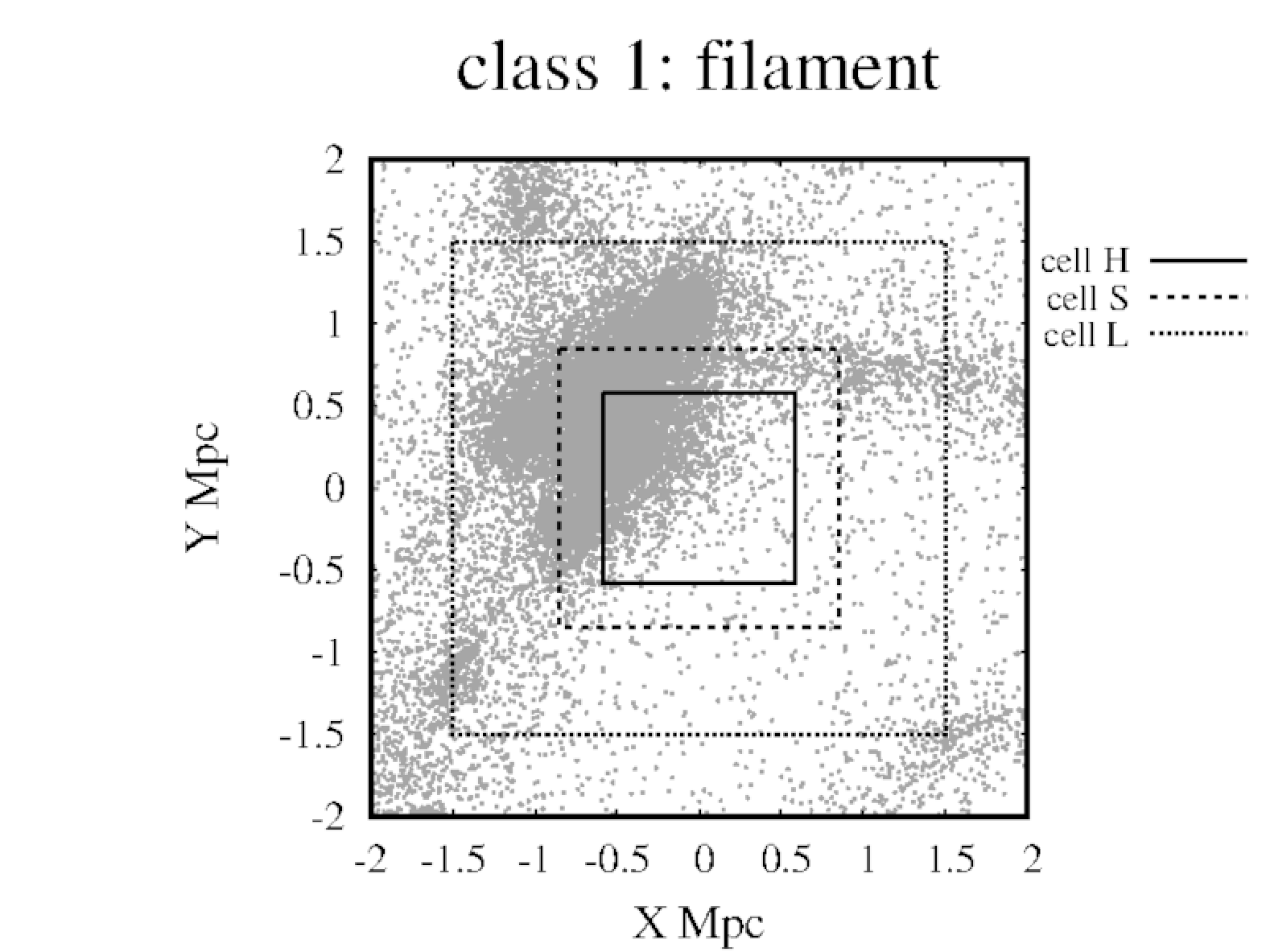} & \includegraphics[width=2.7 in]{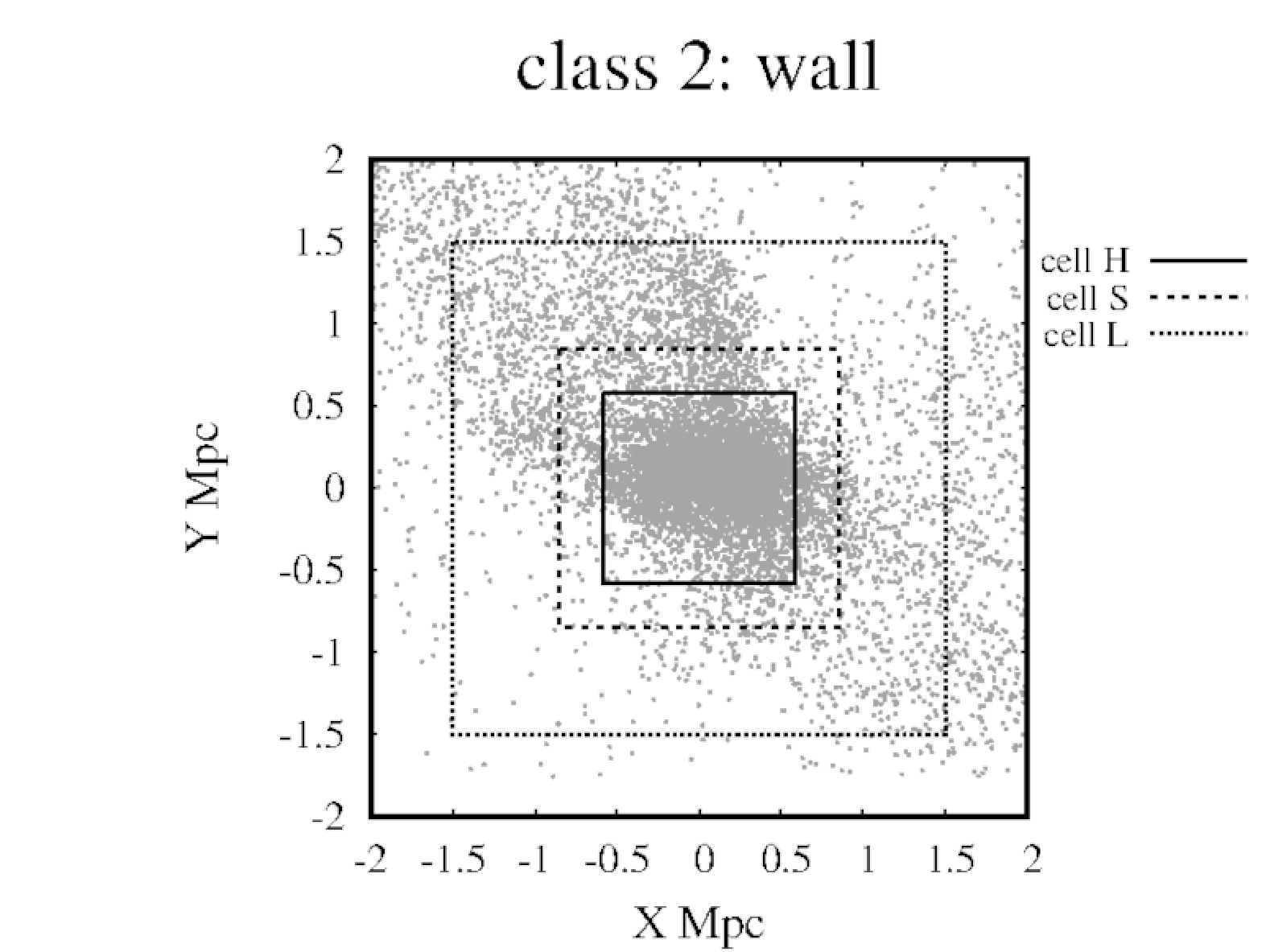} \\
\end{tabular}
\caption{\label{VisCelda} We show two examples of matter structures, in the left panel of class 1 (filament) and the right panel of 
class 2 (wall). We do not show an example of class 3. The different numbers of simulation particles that are captured by the two different 
visualisation cells (types L and S) can be compared with the size of the calculation cell (of type H).}
\end{center}
\end{figure}

To create a plot, we change the origin of the coordinates to the centre of the cell so that the plots are normalised with respect to the side 
length $\Delta_L$. In Fig. \ref{VisCelda} we show examples of class 1 (in the left panel) and class 2 (in the right panel) matter 
structures and compare the size of the three cell types. To summarise, we use a cubic cell of type H to compute the number of positive 
eigenvalues of the Hessian, which is shown as the smallest in Fig. \ref{VisCelda}. Visualisations of cells of type S and type L are also 
shown.

As already mentioned in Section \ref{sec:int}, due to the three-dimensional nature of the cosmic web, it is almost impossible to distinguish 
the different matter structures using only a two-dimensional Cartesian projection. For example, it would not be possible to 
distinguish between a filament and the projection of a sheet. For this reason, we have created an additional model in which the set of 
plots fed to the neural network consist of three projections onto the Cartesian axes; i.e. for each cubic cell we create plots 
of the XY , ZX and ZY projections. For this model we use the label 'LT', as can be seen in Table \ref{tab:numeroimagenes}.  

\begin{figure}
\begin{center}
\begin{tabular}{ccc}
\includegraphics[width=2.0 in]{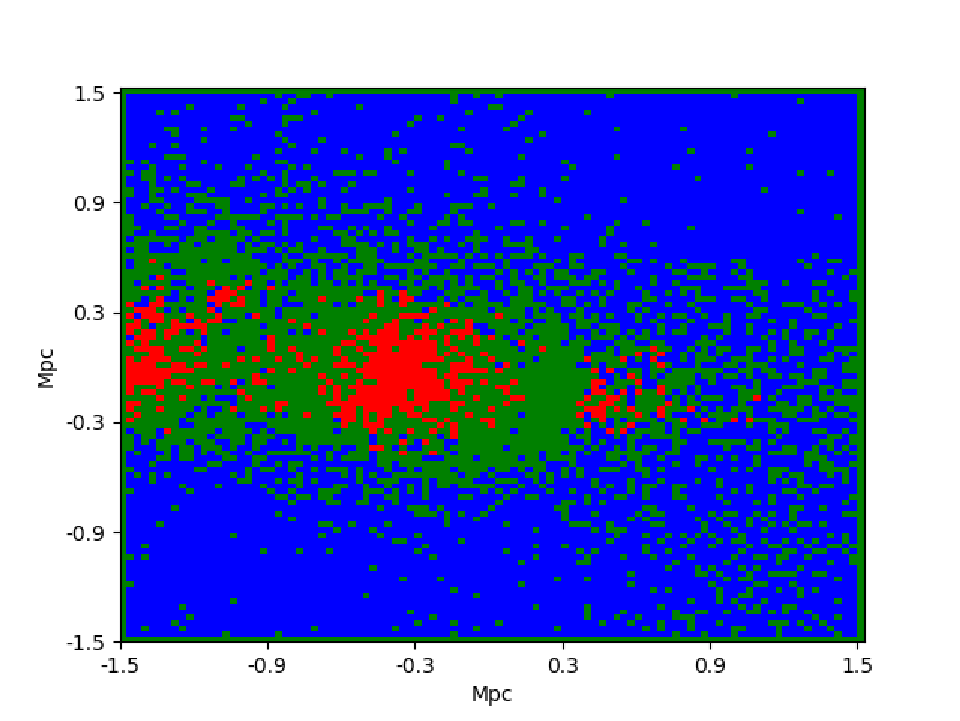} & \includegraphics[width=2.0 in]{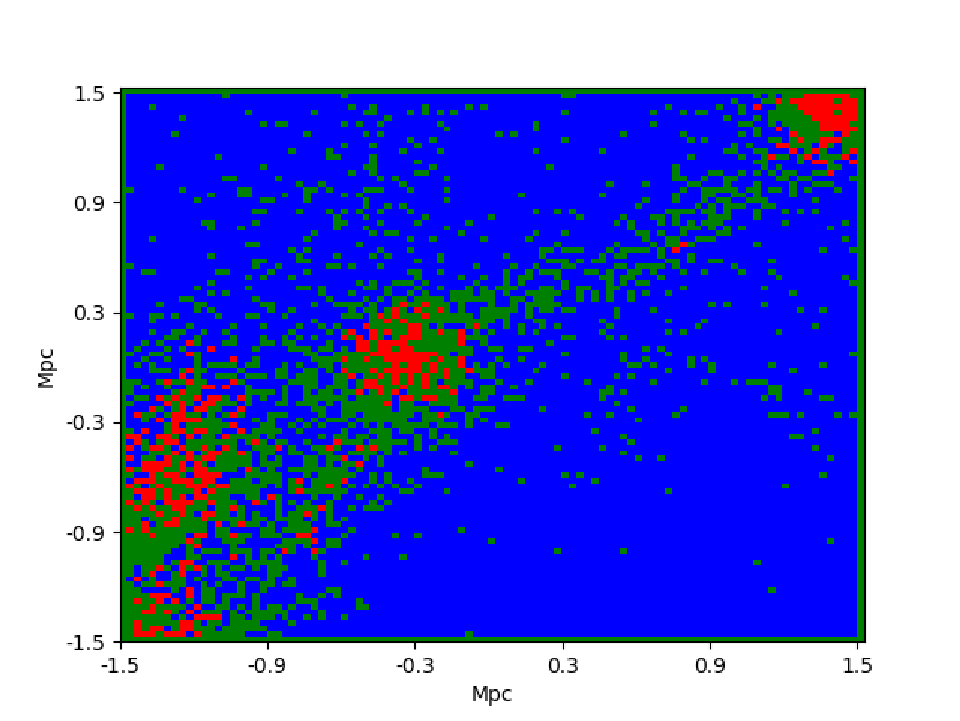} & 
\includegraphics[width=2.0 in]{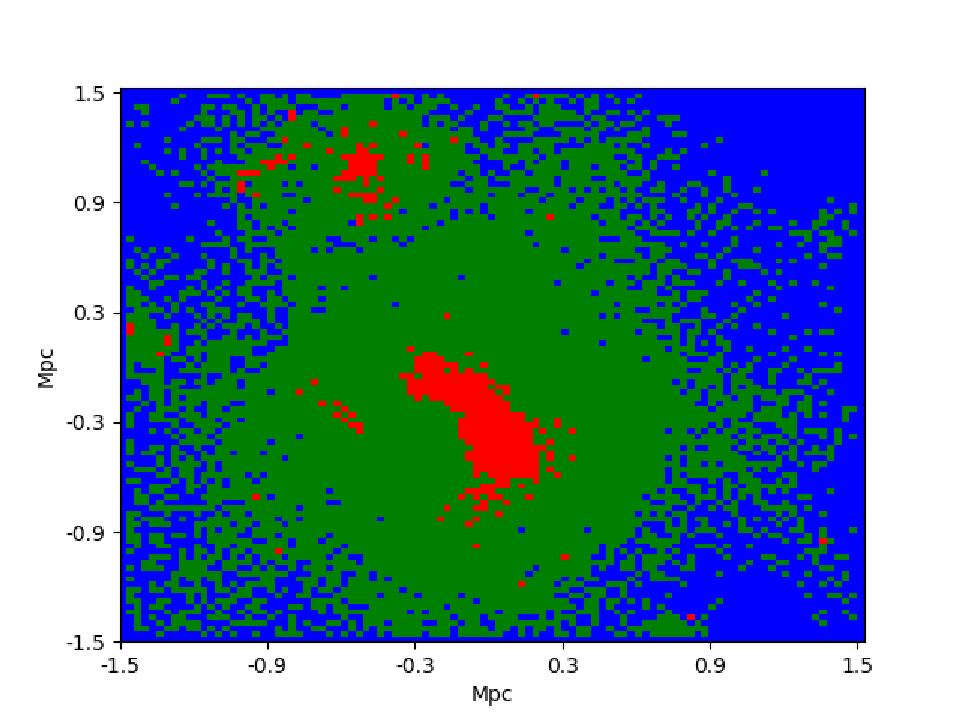}\\
\includegraphics[width=2.0 in]{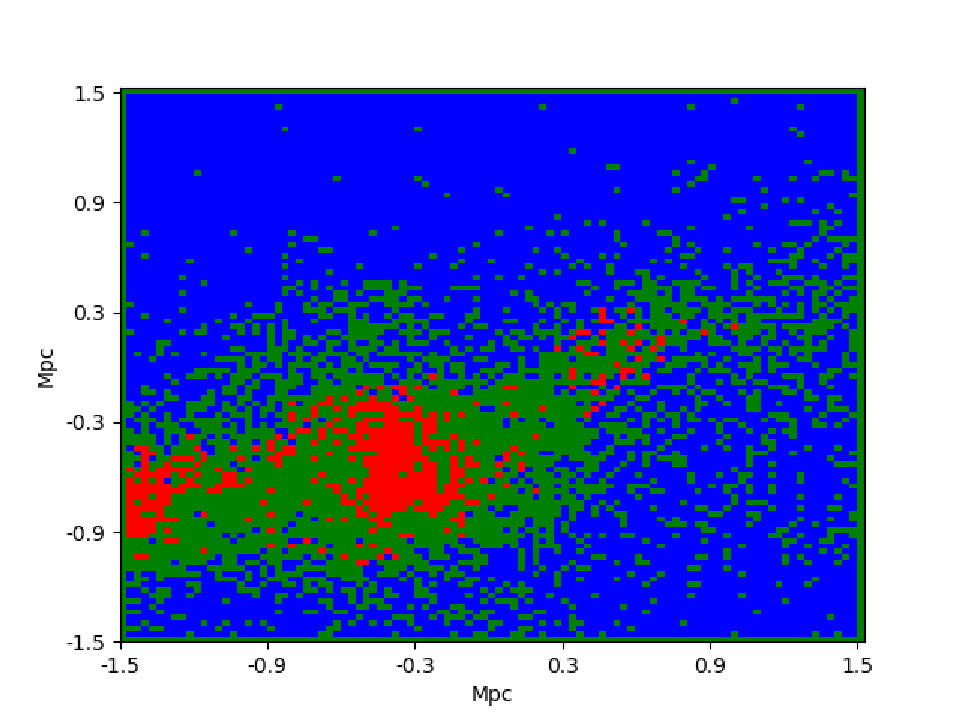} & \includegraphics[width=2.0 in]{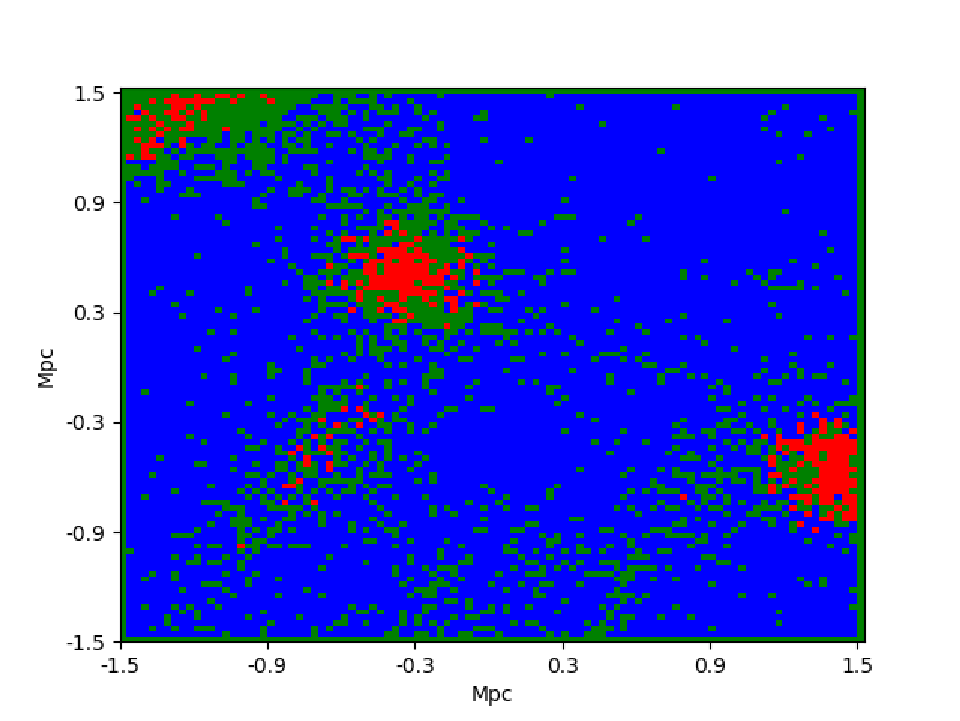} & 
\includegraphics[width=2.0 in]{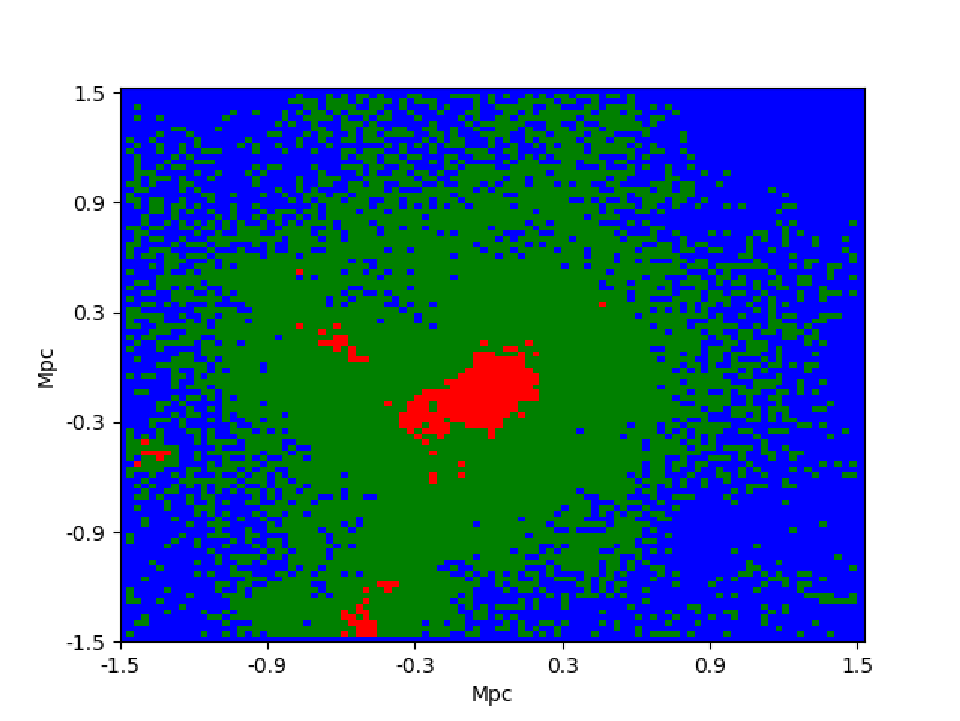}\\
\includegraphics[width=2.0 in]{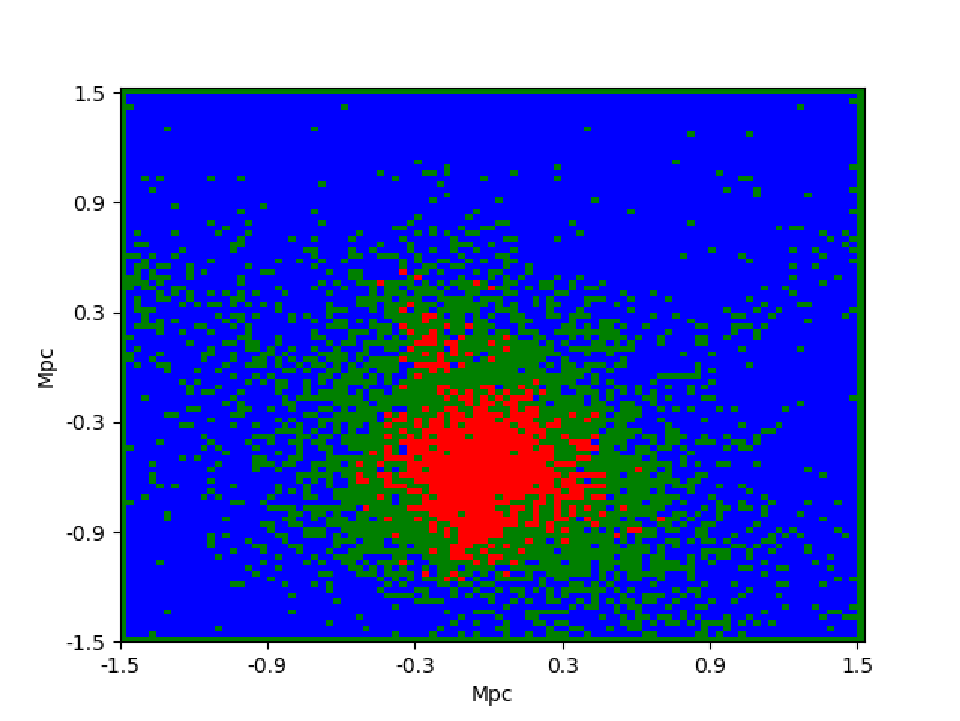} & \includegraphics[width=2.0 in]{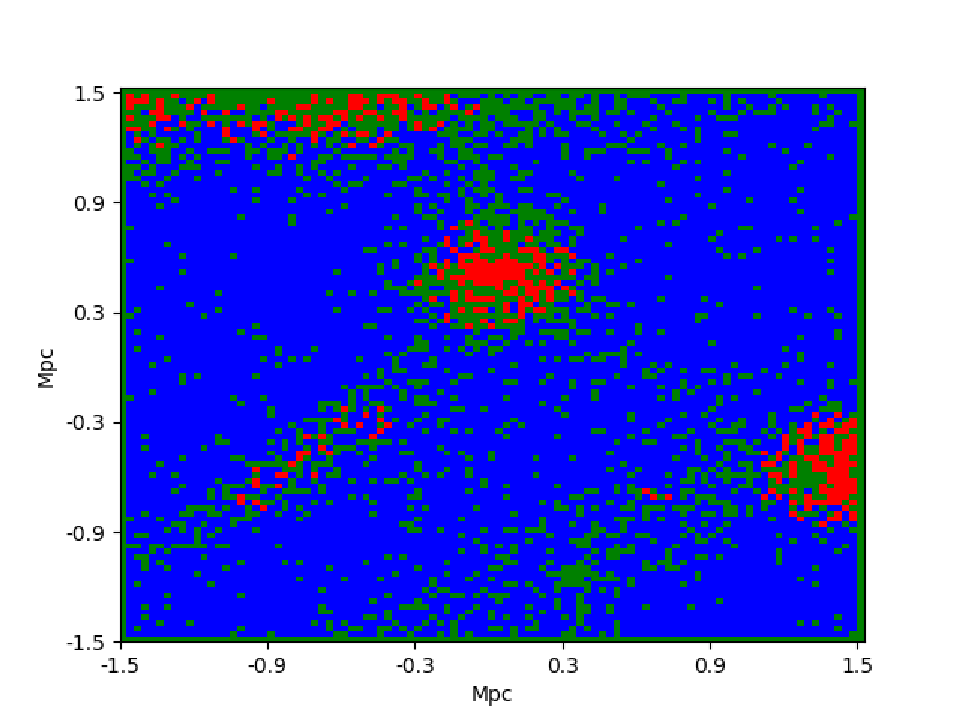} & 
\includegraphics[width=2.0 in]{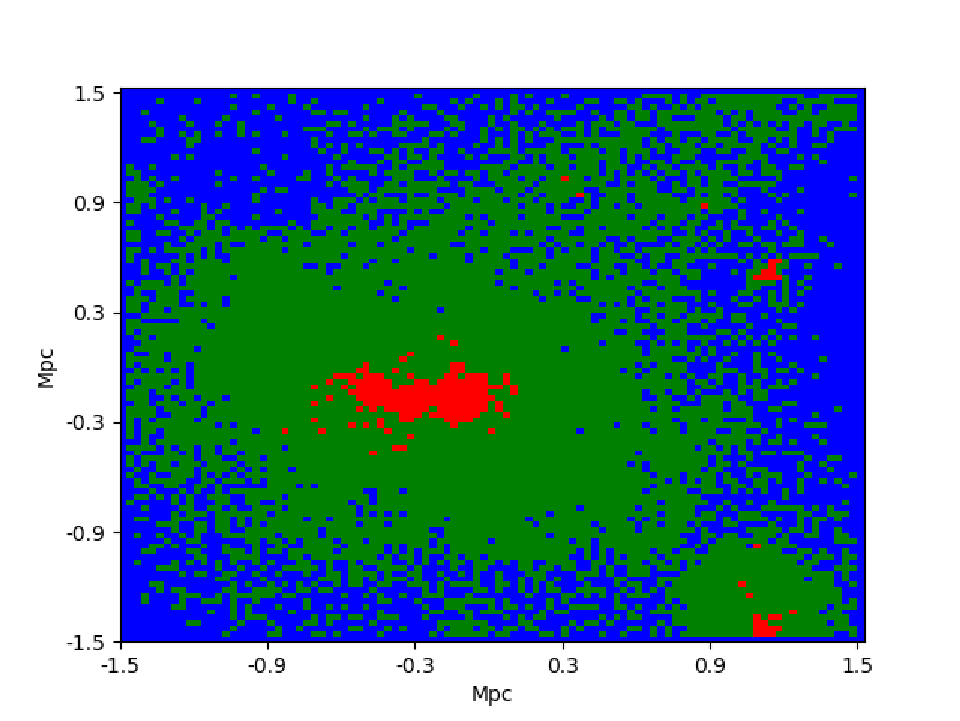}\\
\end{tabular}
\caption{\label{EigenSAN} Examples of isodensity plots were generated for cells with different numbers of positive eigenvalues. The vertical 
and horizontal axes of these plots indicate the length 
in Mpc. For model L, with a width of 578 px and a height of 434 px, the axes vary from -1.5 to 1.5, and we show the projections on the XY 
axes (top line), the ZX axes (second line from top to bottom), and the ZY axes (third line from top to bottom). For model S, with a 
width of 640 px and a height of 480 px, the XY axes vary from -0.85 to 0.85 (not shown in this figure ). Plots are shown with only one positive 
eigenvalue (left column), two positive eigenvalues (middle column) and three positive eigenvalues (right column).}
\end{center}
\end{figure}

In Figure \ref{EigenSAN} we show examples of the training plots. In the left column of Figure \ref{EigenSAN} we show 
the XY (top row), ZX (second row) and ZY (bottom row) projections of a matter structure that could be identified as a 
filament (with only one positive eigenvalue). In the middle column, we show (from top to bottom) the XY, ZX and ZY projections of 
an object that could be identified as a wall (with two positive eigenvalues). In the right column, we show the projections of an object 
that could be identified as a cluster (with three positive eigenvalues). We emphasise that the columns in Figure \ref{EigenSAN} are 
not connected, but represent different matter structures.

The plots were created with the same Python code, based on the plot library Matplotlib. The basic 
features are the same for all plots, for example, they have a resolution of 100x100 dots per inch (dpi). 
The plots for model L have a width of 578 pixels and a height of 434 pixels, while the plots for model S have a 
width of 640 and a height of 480 px. For both model L and model S, the original number of plots was the same, as we 
mentioned in Subsection \ref{subs:cells}.

However, for each plot, we created several additional plots by rotating the original plot, with the 
rotation angle randomly determined with respect to the coordinates centre of each cubic cell. In this way, we 
increased the number of plots available for the training and validation sets. It should be noted that the number 
of rotated plots per original plot is different for each model. For this reason, the number of plots is 
different, as can be seen in Table \ref{tab:numeroimagenes}. 

\begin{figure}
\begin{center}
\includegraphics[width=4.7 in]{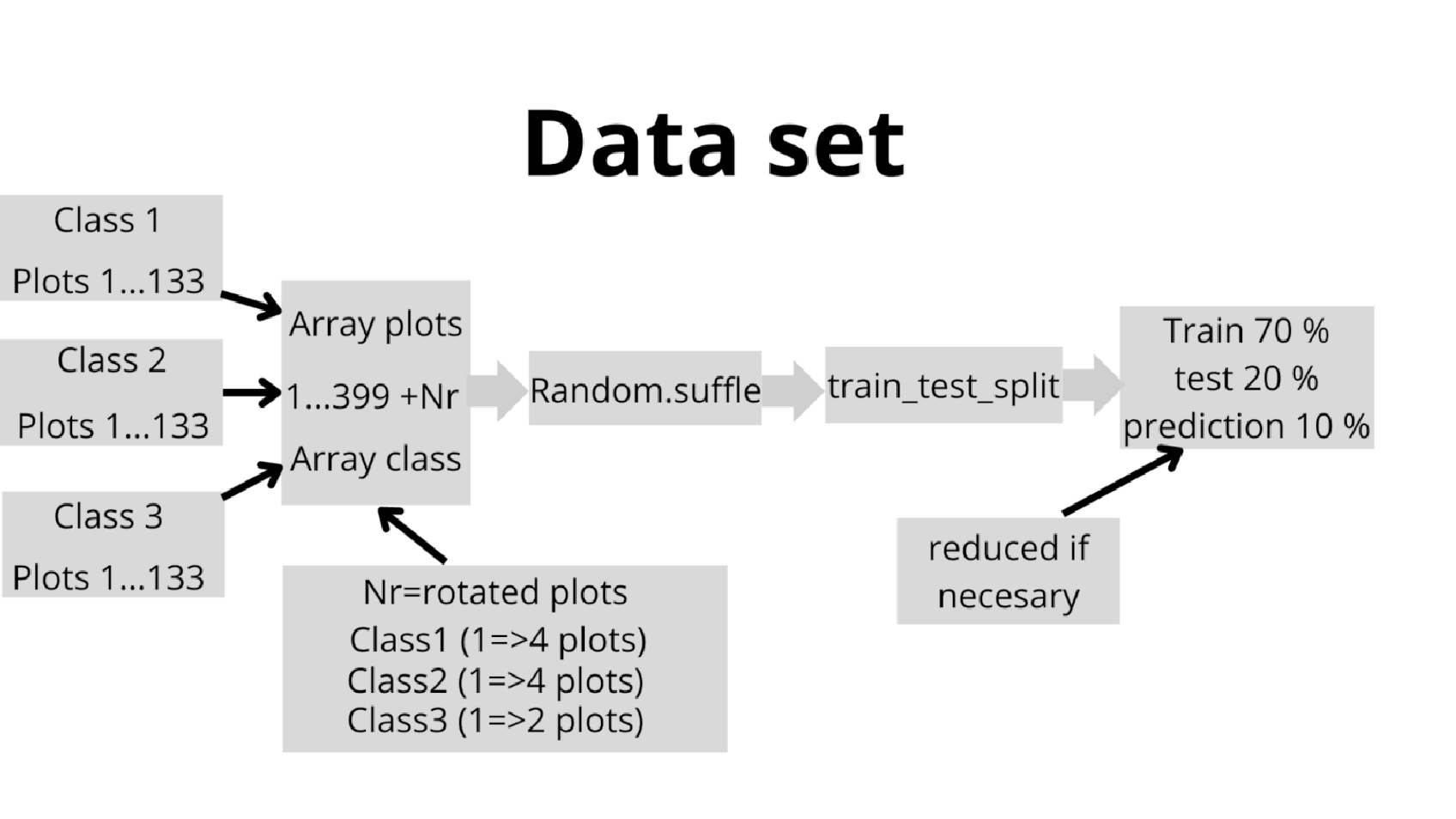} 
\caption{\label{DatosConjunto} Schematic diagram of the generation process of the data set.}
\end{center}
\end{figure}

As we mentioned in Subsection \ref{subs:cells}, the cell catalogue consists of 14515 cells in 
class 3, 145 cells in class 2 and finally 133 cells in class 1. We can create a plot and rotations of this plot 
from each cell catalogue. It must be emphasised 
that we will not use all the cells in class 3, in order to avoid an imbalance in the number of samples 
with respect to the classes. For this reason, we have only considered a subset of class 3 so that 
the number of plot samples is similar for all classes; see Table \ref{tab:numeroimagenes} and Section 
\ref{sec:results}. In Figure \ref{DatosConjunto} we show a schematic diagram of the generation process of the dataset.

To take advantage of the Table \ref{tab:numeroimagenes}, we have included some numbers that we 
will clarify in the Subsection \ref{subs:network} and in the Section \ref{sec:results}, where we 
will present the CNN used to classify the plots and the results\footnote{Rows 10-13 show the results generated during 
the code execution that used the validation data. Line 14 shows the results generated with 
the prediction data}.

\begin{table}[ph]
\caption{The number of images and some results of the CNN models.}
{\begin{tabular}{|c|c|c|c|c|c|c|} \hline
 row &                          & Model L         &  Model S       &  Model LS        & Model LT           \\
1     &  Class 1               &   240            &    360            &   610                   &     600          \\
2     &  Class 2               &   240            &    360            &   662                   &     600           \\
3     &  Class 3               &   217            &    347            &   544                   &    300             \\
4     &  training plots      &   557            &    853            &  1452                   &    1200             \\
5     &  validation plots    &   140            &    214           &   364                    &     300           \\
6     &  prediction plots   &    90              &     90           &   180                    &      180             \\
7     &  total plots           &   697             &   1067         &  1816                    &    1500            \\
8     &  Total parameters &  55948771     &  69973475    &  69973475              & 69973475         \\
9     &  Memory             &  214 MB         &   266 MB      &  266 MB                 &    266                \\
10    &  correct labels     &  100              &   175            &   256                     &      201               \\
11    &  incorrect labels   &   40               &   39             &   108                    &      99               \\
12    &  test accuracy      &   0.71            &  0.81            &   0.70                  &      0.67              \\
13    &  test loss             &   0.49             &  0.48            &   0.58                  &     0.63              \\
14    &  true positives     & 0.23,0.16,0.3   & 0.26,0.1, 0.33 &  0.11, 0.11, 0.16  &   0.28,0.1,0.18    \\
15    &  $\Delta_L$         &   1.5 Mpc       &  0.85 Mpc       &  1.5 Mpc and 0.85 Mpc &        1.5 Mpc       \\
\hline
\end{tabular} }
\label{tab:numeroimagenes}
\end{table}    

\subsection{The CNN}
\label{subs:network}

The CNN design of this paper was motivated by many successful models, including, among others, 
the simple MNIST convnet \citet{chollet}; in particular, the CNN for classifying sports described 
by \citet{bagnato} and the CNN described by \citet{bagnato} for classifying dogs and cats. In Table \ref{tab:cnnmodel} we 
describe the CNN model considered in this paper. 

\begin{table}[ph]
\caption{Layers of the CNN, model sequential-5.}
{\begin{tabular}{|c|c|c|} \hline
Layer (type)                        & Output Shape              & Number of Parameters  \\
conv2d$_{14}$ (Conv2D)                & (None, 576, 432, 32)      & 1184                  \\            
max-pooling2d$_{14}$ (MaxPooling2D)   & (None, 288, 216, 32)      &     0                 \\
conv2d$_{15}$ (Conv2D)                & (None, 286, 214, 64)      &  18496                 \\     
max-pooling2d$_{15}$ (MaxPooling2D)     & (None, 143, 107, 64)      &     0                 \\         
conv2d$_{16}$ (Conv2D)                & (None, 141, 105, 128)     &  73856                \\     
max-pooling2d$_{16}$ (MaxPooling2D)     & (None, 70, 52, 128)       &     0                   \\        
conv2d$_{17}$ (Conv2D)                & (None, 68, 50, 128)       &   147584              \\    
max-pooling2d$_{17}$ (MaxPooling2D)   & (None, 34, 25, 128)       &     0                  \\        
flatten$_5$ (Flatten)               & (None, 108800)            &     0                   \\         
dense$_{10}$ (Dense)                  & (None, 512)               &   55706112              \\ 
dense$_{11}$ (Dense)                  & (None, 3)                 &    1539                 \\     
\hline
\end{tabular} }
\label{tab:cnnmodel}
\end{table}

\begin{figure}
\begin{center}
\includegraphics[width=4.7 in]{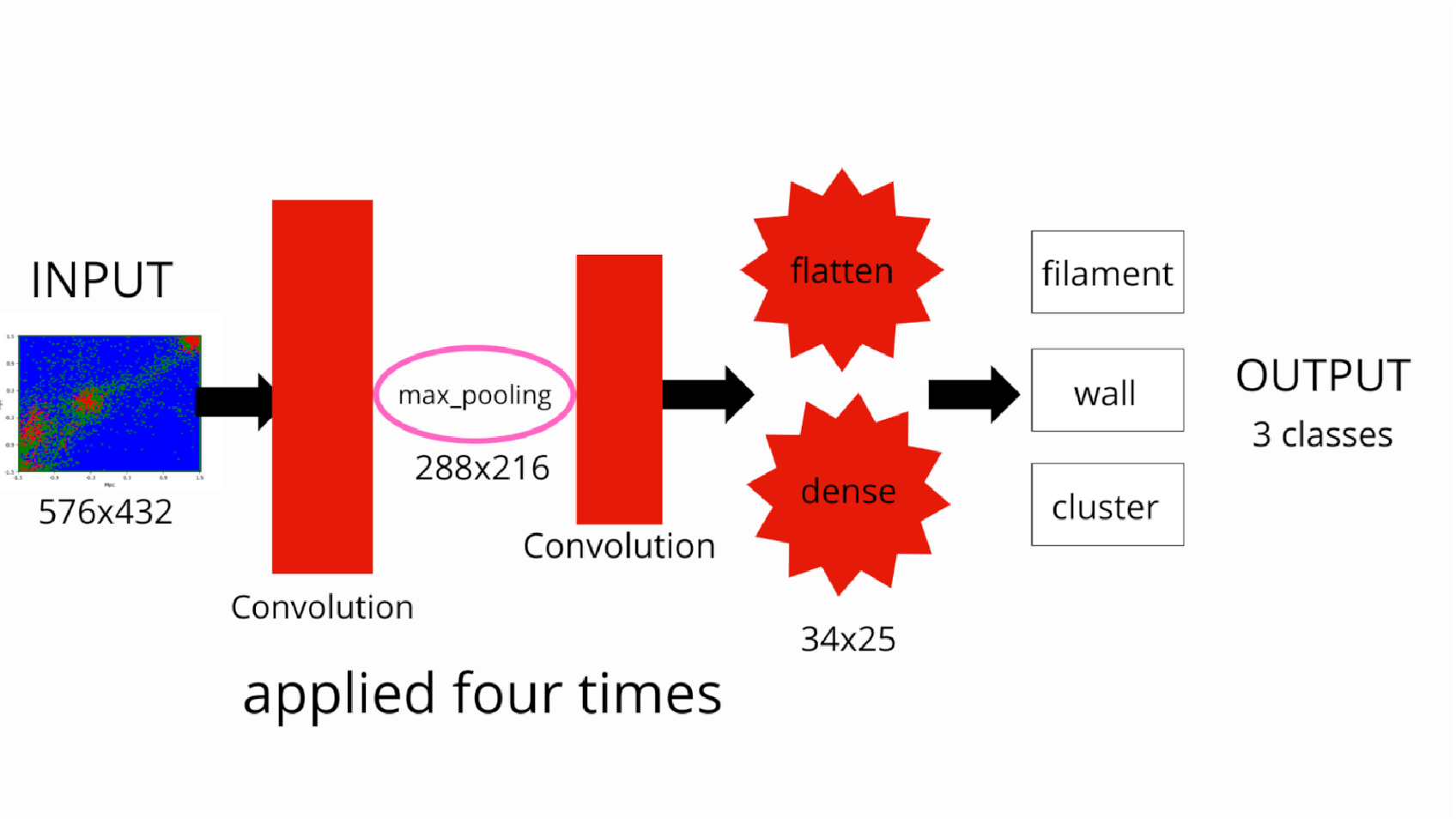} 
\caption{\label{Convolution} Schematic diagram of the CNN.}
\end{center}
\end{figure}

In Figure \ref{Convolution} we show a schematic representation of the CNN with the main layers and their effects on the 
size of the plot (in pixels).  In general, a CNN has two main stages: one for feature extraction and the other for 
classification. In the extraction stage, the input plot is passed to a convolutional layer to create a feature map of the plot, the 
size of which is then reduced by a pooling process. This step can be applied several times to the same input plot. 

In the classification phase, the feature matrix is converted into a vector and the values of the parameters available in the CNN are 
determined so that the results match the class of the plot. To compare the performance and quality of the results of 
the CNN described in this Section \ref{subs:network}, we consider an alternative CNN in the Appendix \ref{app:network}. However, other 
models can also be tested to explore different CNN options. 

It should be noted that in both CNNs, the densest layers are those that offer the largest number of adjustable parameters. In the 
alternative CNN shown in Appendix \ref{app:network}, only the dense layer generates parameters. The other layers have other 
goals, such as creating a character map, or reducing the number of pixels, but the dense layers must establish the connection 
between all the neurons of the previous layer and the next layer, which is why they must have a parameter for each connection. 

\section{Results}
\label{sec:results}

We apply the CNN described in Subsection \ref{subs:network} to several sets of plots. The CNN is trained during different 
epochs for all models to determine the best values of the free parameters to assign each plot to the target class. The accuracy and loss 
metrics are two of the best known metrics for determining the performance 
of a CNN. Accuracy is expressed as a percentage, i.e. it 
indicates the proportion of plots whose classes were correctly predicted by the CNN. The loss metric, also 
known as the cost or error function, measures the failure in predicting the correct class for a plot with 
respect to the target class and is expressed as a real number.  

\subsection{Report on training and validation}
\label{sec:report}

\begin{figure}
\begin{center}
\begin{tabular}{cc}
\includegraphics[width=2.5 in]{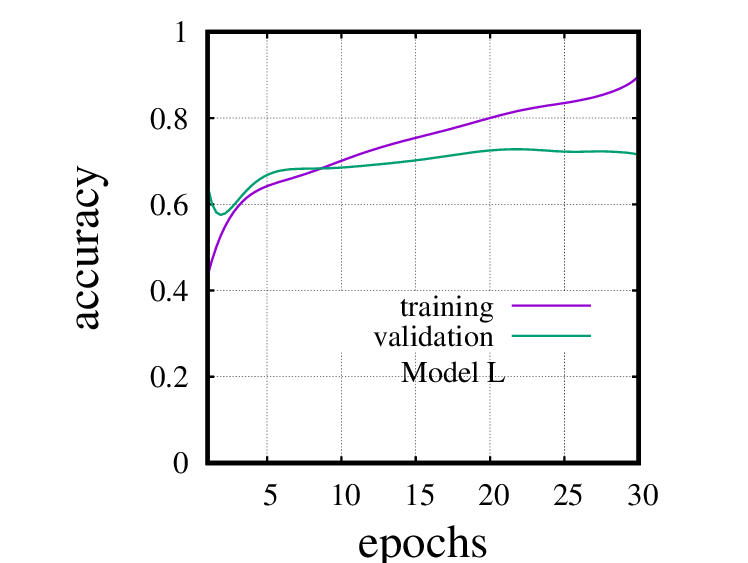} & \includegraphics[width=2.5 in]{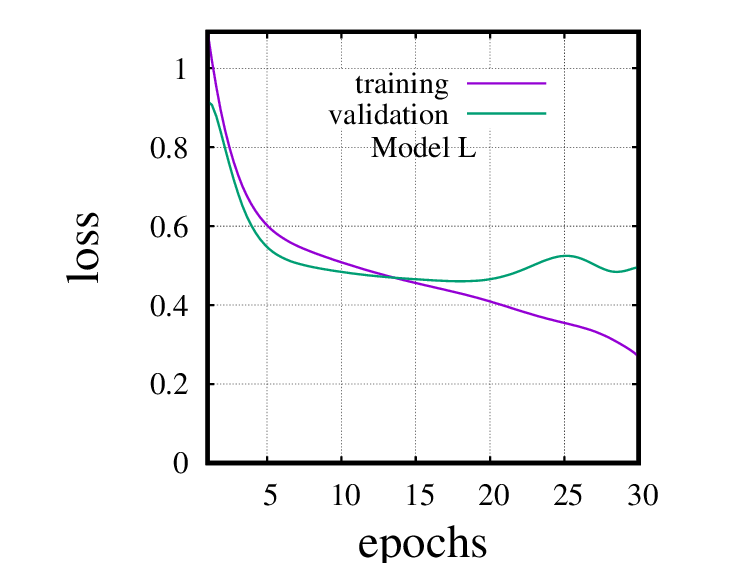} \\
\includegraphics[width=2.5 in]{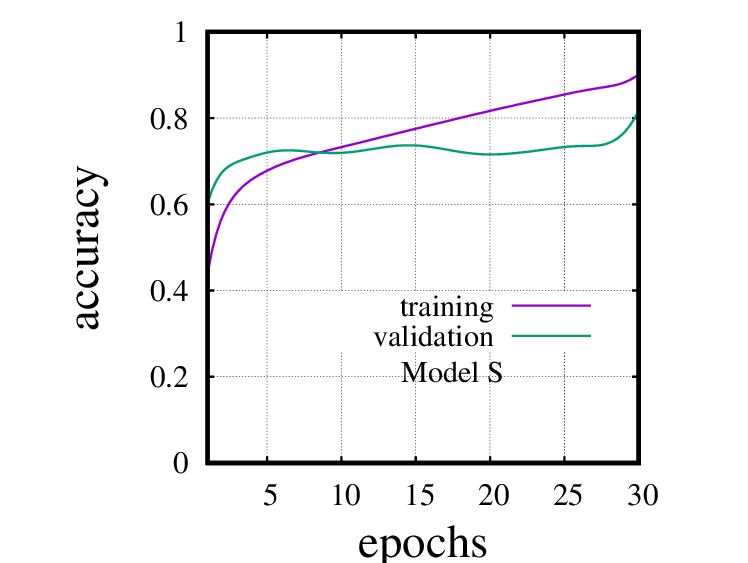} & \includegraphics[width=2.5 in]{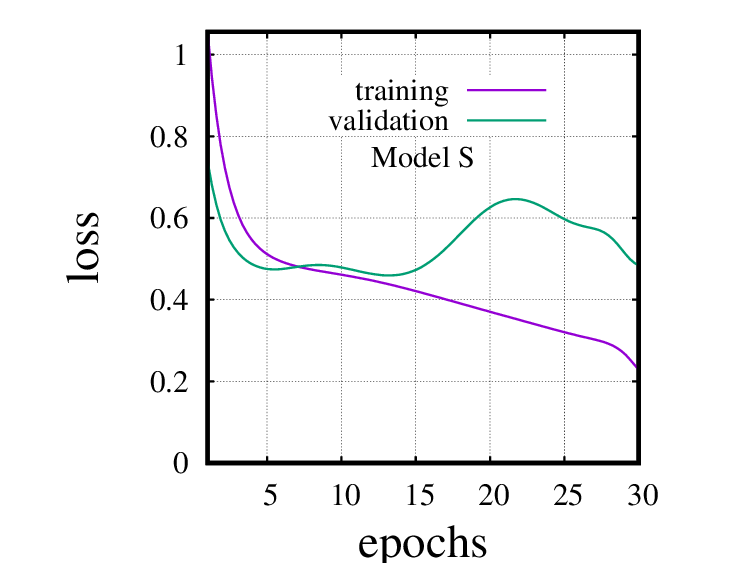} \\
\includegraphics[width=2.5 in]{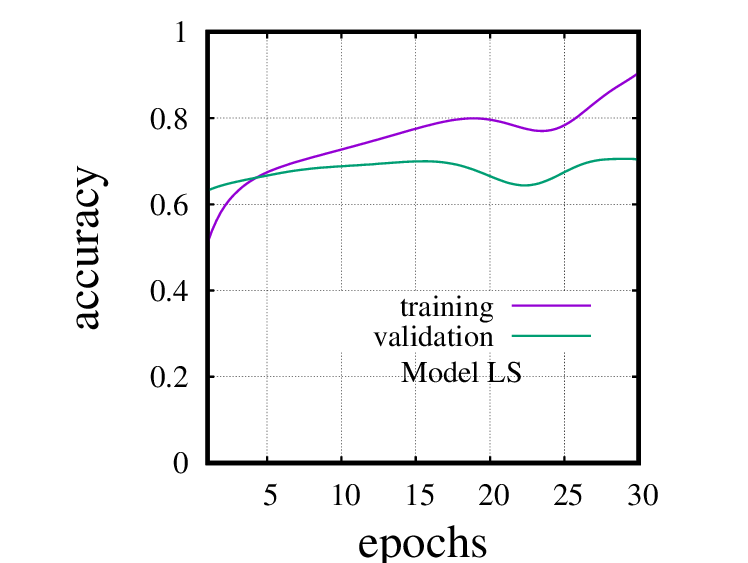} & \includegraphics[width=2.5 in]{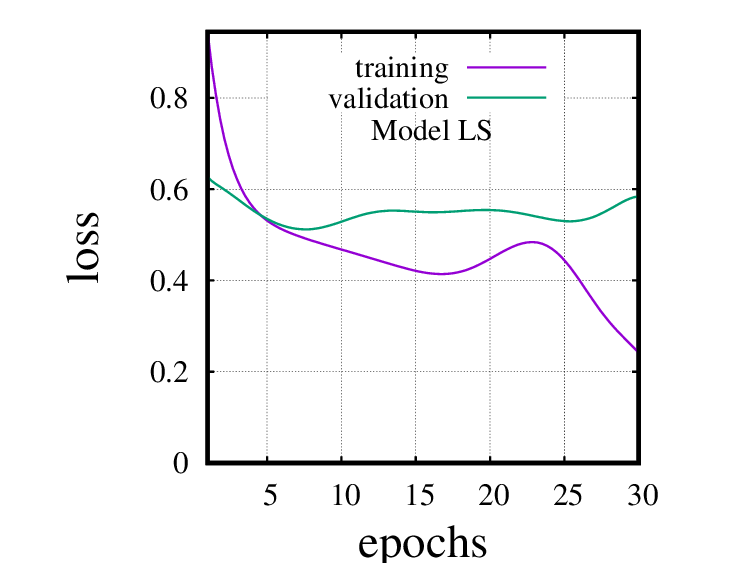} \\
\includegraphics[width=2.5 in]{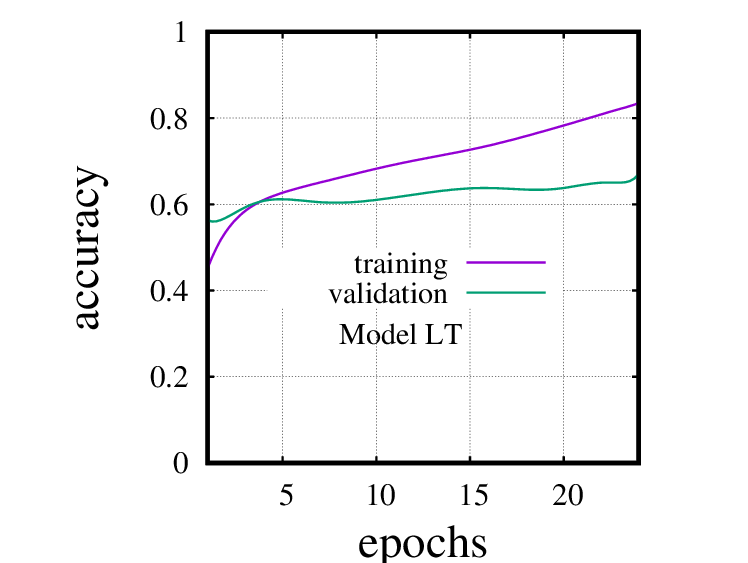} & \includegraphics[width=2.5 in]{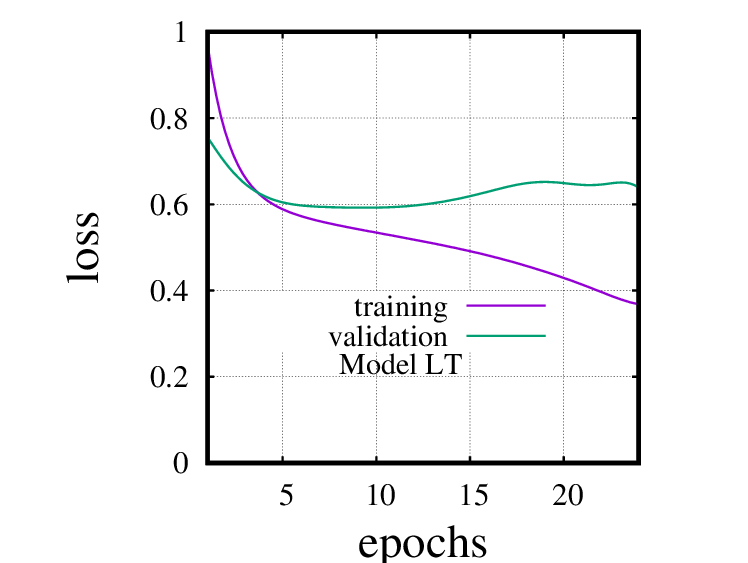} \\
\end{tabular}
\caption{\label{networkperformance} Accuracy and loss history in terms of the epochs, for
model L (top line); model S (second line from top), model LS ( third line from top ) and model LT (bottom line). }
\end{center}
\end{figure}

In Fig.\ref{networkperformance} we show the accuracy and loss metrics in relation to the epochs. To determine 
the accuracy and loss metrics, the code uses the validation plots. The curves of all models 
show the expected behaviour: the accuracy and validation curves for the training plots increase over the epochs. The same 
observation can be made for the loss curves, i.e. the loss curves for the training and validation plots 
decrease as the number of epochs increases.   

\begin{table}[ph]
\caption{Report for model L.}
{\begin{tabular}{|c|c|c|c|c|} \hline
       --      &  precision  &  recall & f1-score  & support \\
     Class 1   &    0.56     &   0.50  &    0.53   &     44  \\
     Class 2   &    0.52     &   0.56  &    0.54   &     41  \\
     Class 3   &    0.96     &   1.0   &    0.98   &     55  \\
    accuracy   &     --      &   --    &    0.71   &    140  \\
   macro avg   &    0.68     &   0.69  &    0.68   &    140  \\
\hline
\end{tabular} }
\label{tab:f1scoreL}
\end{table}

\begin{table}[ph]
\caption{Report for model S.}
{\begin{tabular}{|c|c|c|c|c|} \hline
       --      &  precision  &  recall & f1-score  & support \\
     Class 1   &    0.73     &   0.52   &   0.61    &    58   \\
     Class 2   &    0.71     &   0.86   &   0.78    &    79   \\
     Class 3   &    1.0      &   1.0    &   1.0     &    77  \\
    accuracy   &    --       &    --    &   0.82    &    214  \\
   macro avg   &    0.81     &   0.79   &   0.79    &    214  \\
\hline
\end{tabular} }
\label{tab:f1scoreS}
\end{table}

\begin{table}[ph]
\caption{Report for model LS.}
{\begin{tabular}{|c|c|c|c|c|} \hline
       --      &  precision  &  recall & f1-score  & support \\
     Class 1   &    0.61     &   0.33   &   0.42    &   120   \\
     Class 2   &    0.55     &   0.78   &   0.65    &   123   \\
     Class 3   &    0.95     &   1.0    &   0.98    &   121  \\
    accuracy   &    --       &    --    &   0.70    &   364  \\
   macro avg   &    0.71     &   0.70   &   0.68    &   364 \\
\hline
\end{tabular} }
\label{tab:f1scoreLS}
\end{table}

\begin{table}[ph]
\caption{Report for model LT.}
{\begin{tabular}{|c|c|c|c|c|} \hline
       --      &  precision  &  recall & f1-score  & support \\
     Class 1   &    0.61     &   0.71   &   0.65    &   130   \\
     Class 2   &    0.59     &   0.46   &   0.52    &   114   \\
     Class 3   &    0.97      &   1.0    &   0.98   &    56  \\
    accuracy   &    --       &    --    &   0.67    &   300  \\
   macro avg   &    0.72     &   0.72   &   0.72    &   300 \\
\hline
\end{tabular} }
\label{tab:f1scoreLT}
\end{table}

To better examine the performance of the CNN, we need to go beyond accuracy and loss metrics. The f1-score 
is another metric that can be used in addition to the two simpler performance metrics 
mentioned above, namely the accuracy and loss. To explain the f1-score metric, we first introduce 
the concepts of precision and recall. Precision is defined as the ratio between the number of 
correct positives prediction and the sum of the correct positive predictions and the number of false positives predictions, i.e.
precision measures how many of the CNN's positive predictions were correct. 
Recall is defined as the ratio between the number of true positives and the sum of the number of 
true positives and the number of false negatives, i.e. recall measures how 
many positives predictions the CNN found out of all positives.

The f1-score combines the metrics precision and recall into a single metric. The f1-score 
is defined as the ratio between 2 times the precision times the recall and the sum of precision 
and recall, i.e. the f1-score is an average of precision and recall; see 
Tables \ref{tab:f1scoreL}, \ref{tab:f1scoreS}, \ref{tab:f1scoreLS} and \ref{tab:f1scoreLT}.
The macro-average values for precision and recall are calculated as the average of the individual precision and 
recall values of the individual classes.  

For model L, we can see that for classes 1 and 2, which identify filaments and walls, the CNN 
recognises these classes with an f1-scores of only 0.53 and 0.54, respectively; see Table \ref{tab:f1scoreL}. For the S model, 
these values are 0.61 and 0.78 respectively; see Table \ref{tab:f1scoreS}. For the LS model, these 
values are 0.42 and 0.65 respectively; see Table \ref{tab:f1scoreLS}. For the LT model, these values are 0.65 
and 0.52 respectively, see Table \ref{tab:f1scoreLT}. Based on the f1- metric, model S and model LT deliver 
the best results. The worst results are achieved by the LS model.

If we look at the results of all models, we can summarise them as follows:
the class 3 objects that identify clusters, are recognised very well. This is the 
expected behaviour: recognising filaments and walls is much more difficult than 
the recognising clusters.  It seems that the smallest cell size provides a better performance in the detection of 
filaments and walls.

\subsection{Report on prediction}
\label{sec:reportpredi}

\begin{figure}
\begin{center}
\begin{tabular}{cc}
\includegraphics[width=2.5 in]{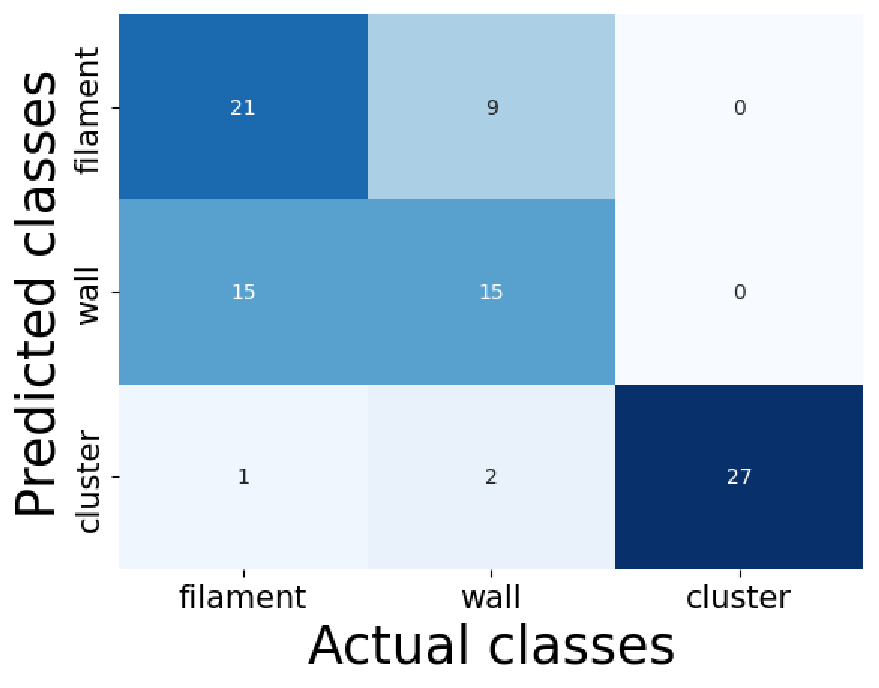} & \includegraphics[width=2.5 in]{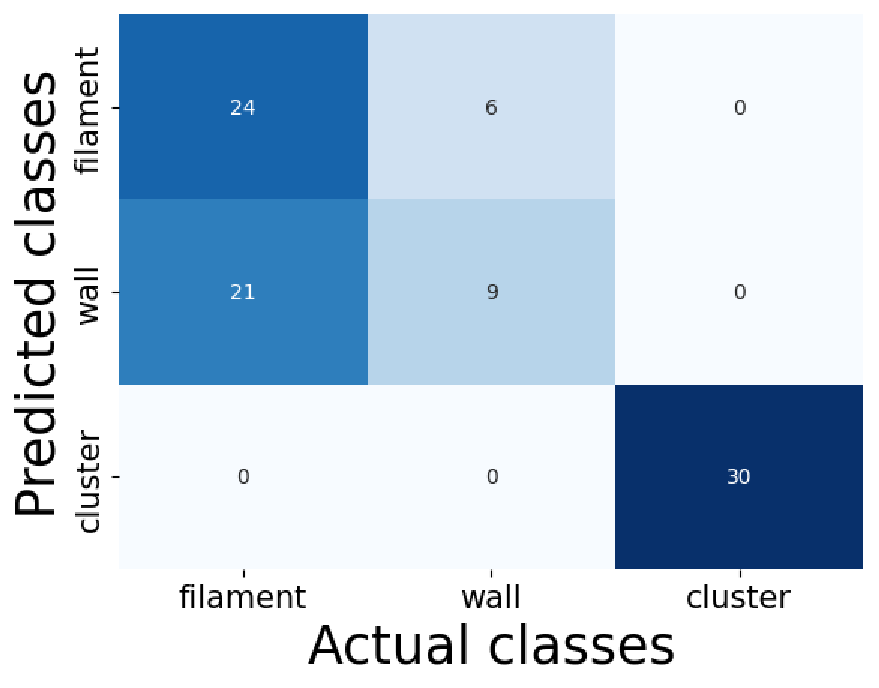}\\
\includegraphics[width=2.5 in]{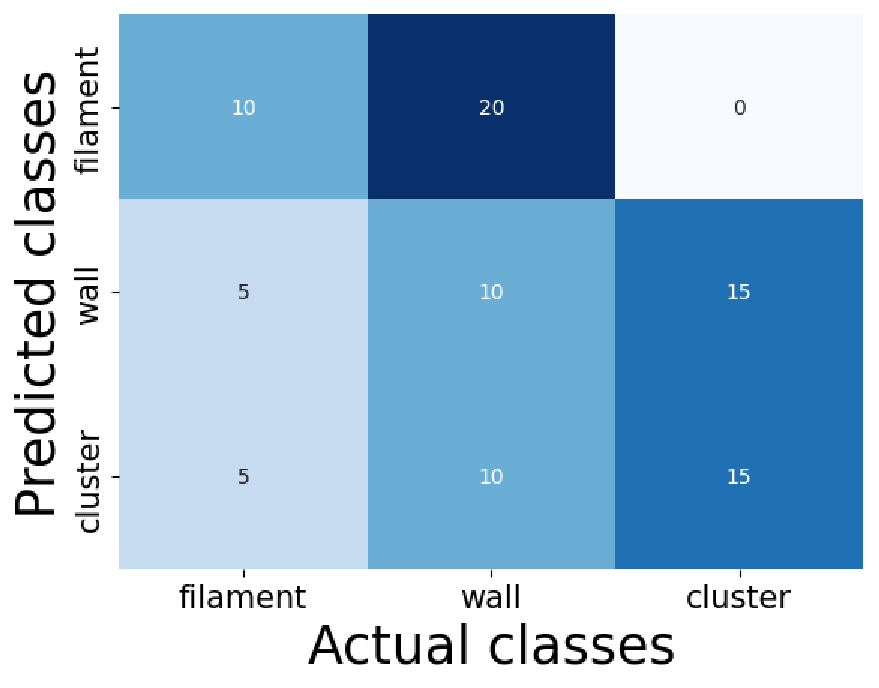}  & \includegraphics[width=2.5 in]{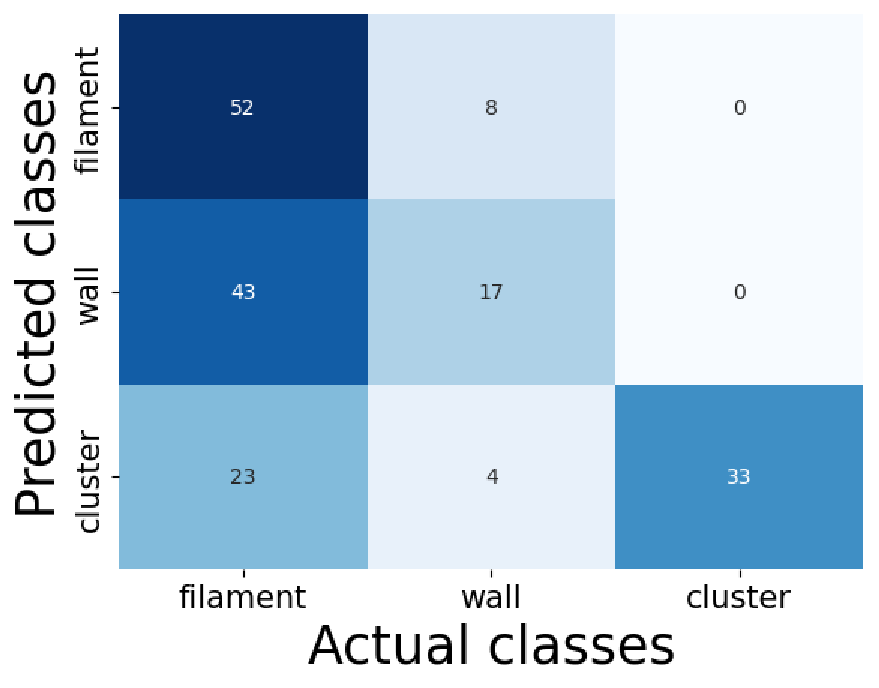}\\
\end{tabular}
\caption{\label{MiclasspredictionL} Confusion matrix for unseen plots. Model L (top left); model S (top right); 
model LS (bottom left) and model LT (bottom right). }
\end{center}
\end{figure}

Next, we use plots (which are not included in the training or validation sets) to see 
if the CNN can correctly predict the classes of these unseen plots. The results are shown in 
Fig. \ref{MiclasspredictionL} in the form of a confusion matrix and in the Tables \ref{tab:f1scoreLpred}, 
\ref{tab:f1scoreSpred}, \ref{tab:f1scoreLSpred} and \ref{tab:f1scoreLTpred}. Let us now present the results.  

For model L, the CNN returned 63 correct labels and 
27 incorrect labels, which means that the proportion of true positives for classes 1,2 and 3 is 
0.23, 0.16 and 0.3, respectively. From these results, it can be concluded that the CNN has a 
global accuracy of 69 percent ( the sum of all these proportions ). For model S, the CNN 
provided 63 correct labels and 27 incorrect labels. The proportions of true positives are 
0.26, 0.1 and 0.33, for classes 1, 2 and 3 respectively. The sum of all these proportions 
is 69 percent. These values are also shown in Table \ref{tab:numeroimagenes}.

Model LS was an attempt to combine model L and model S, i.e. all the plots of individual models were combined 
in model LS. However, the size of the array was unmanageable due to the high memory 
requirements. For this reason, we only use the first half of the plots of each model; see 
Table \ref{tab:numeroimagenes}. As we mentioned in the caption of Fig.\ref{EigenSAN}, the 
sizes of the plots in model L and model S are different. In order to obtain plots of uniform size 
for model LS, i.e. plots with the same width and height, we applied a procedure to scale the plots 
of model L so that they have the same size as the plots of model S. Nevertheless, the length scale of 
the coordinate axes is still different. We believe that this difference does not change 
the results of the neural network.

Using the LS model, we tried to test whether the CNN is sensitive to the different length scales of the 
plots. However, it seems that the results of this model LS change significantly compared to 
model L and model S; see Table \ref{tab:f1scoreLS}. The confusion matrix of model LS has more unseen plots 
than that of model L or model S, as can be seen in the bottom panel of Fig.\ref{MiclasspredictionL}. For model LS, the 
total number of unseen plots was also 90, of which 35 were correctly labelled and 55 were incorrectly labelled. The proportions 
of true positives are 0.1 for class 1, 0.1 for class 2 and 0.16 for class 3. The sum of all these values 
is 0.36. This value is much smaller than the values obtained for model L (0.69) and model S (0.69). 

For the LT model, the proportions of true positives are 0.28, 0.1 and 0.18, respectively. The sum of these proportions is 
0.56. Based on the CNN's ability to make predictions,  the L and S models therefore deliver the best results. The LS model 
delivers the worst results.

\begin{table}[ph]
\caption{Report on prediction for model L.}
{\begin{tabular}{|c|c|c|c|c|} \hline
   --        &  precision  &  recall   & f1-score  &   support \\
class 1         &  0.7          &  0.57     &    0.63    &   37       \\
class 2         &  0.5          &  0.58     &    0.54    &   26    \\
class 3          &  0.9         &  1.0       &    0.94    &  27   \\
accuracy &   0.7         &  0.7       &    0.7      &  0.7   \\
macro avg  &    0.7     &  0.71     &   0.70     &   90  \\
\hline
\end{tabular} }
\label{tab:f1scoreLpred}
\end{table}

\begin{table}[ph]
\caption{Report on prediction for model S.}
{\begin{tabular}{|c|c|c|c|c|} \hline
   --        &  precision  &  recall   & f1-score  &   support \\
class 1         &  0.8         &  0.53     &    0.64    &   45       \\
class 2         &  0.3          &  0.6     &    0.4    &   15    \\
class 3          &  1.0        &  1.0       &    1.0    &  30   \\
accuracy &   0.7         &  0.7       &    0.7      &  0.7   \\
macro avg  &    0.7     &  0.71     &   0.68     &   90  \\
\hline
\end{tabular} }
\label{tab:f1scoreSpred}
\end{table}

\begin{table}[ph]
\caption{Report on prediction for model LS.}
{\begin{tabular}{|c|c|c|c|c|} \hline
   --        &  precision  &  recall   & f1-score  &   support \\
class 1        &  0.33         &  0.5     &    0.4    &   20       \\
class 2         &  0.33          &  0.25     &    0.29    &   40    \\
class 3          &  0.5        &  0.5       &    0.5    &  30   \\
accuracy &   0.38         &  0.38       &    0.38      &  0.38   \\
macro avg  &    0.38     &  0.41     &   0.39     &   90  \\
\hline
\end{tabular} }
\label{tab:f1scoreLSpred}
\end{table}

\begin{table}[ph]
\caption{Report on prediction for model LT.}
{\begin{tabular}{|c|c|c|c|c|} \hline
   --        &  precision  &  recall   & f1-score  &   support \\
class 1         &  0.86         &  0.44     &    0.58   &   118       \\
class 2         &  0.28          &  0.58     &    0.38    &   29    \\
class 3          &  0.55        &  1.0       &    0.70    &  33   \\
accuracy &   0.56        &  0.56       &    0.56      &  0.56   \\
macro avg  &    0.56     &  0.67     &   0.55     &   180  \\
\hline
\end{tabular} }
\label{tab:f1scoreLTpred}
\end{table}

From the f1-scores for the prediction, we can see that the LS model gives the worst results of 0.4, 0.25 and 0.5 for class 1,2 and 3 
matter structures. The L and S models provide better results of 0.63. The LT model provides results at an intermediate level compared with 
to the others models. The L and S models also provide high identification performance for class 3 matter structures, as we have 
seen so far for all models. It should be emphasised that the LS and LT models have lower predictability 
for class 3 matter structures based on this metric. 

\section{Discussion: some concerns}
\label{sec:dis}

The aim of this paper was to use a typical CNN to classify a set of Cartesian slice plots, each of which was 
constructed using the simulation particles located in ( and around) a set of cubic cells ( of type L and S ) that are part of a 
uniform partition of the simulation volume ( cell of type H). The first problem is that the assigned labels may 
not be entirely correct, but they are assumed to be the true target in CNN, which means that we examined 
the results under this assumption. However, other codes might assign different labels 
to the same set of cells, as we explain below.

After we have completed this work, we have found the code TVWEB of \citet{forero}. This code determines a label 
for all grid elements of a uniform partition, regardless of the number of simulation particles 
contained in each grid element. Then, through a smoothing process with a Gaussian kernel, this method obtains 
the density field, which is then transformed into Fourier space to obtain the diagonalization of the 
Hessian matrix.

The output of the TVWEB code provides different numbers than the ones we obtained, namely 113209 cells 
in class 1, 121109 cells in class 2 and 1416 cells in class 3. TVWEB recognises  
many cells in class 1 and class 2, while the number of cells in class 3 is much smaller. As we have shown in 
Subsection \ref{subs:cells}, the number of cells labelled as 
class 3 with our code was 14515 (with three positive eigenvalues), which is much larger than the number of cells 
labelled as class 2 and class 1 ( 145 and 133, respectively) and the number of cells of class 3 found with TVWEB. At this point, it 
must be mentioned that the number of cells with non-positive eigenvalues (including all cells 
discarded from the calculation because the number of simulation particles was below the threshold) found by 
our code was 247351. TVWEB found 26410 cells with non-positive eigenvalues. The total number of cells counted for both codes is 
262144, which takes into account all the cells in the partition.    

\begin{figure}
\begin{center}
\includegraphics[width=5 in]{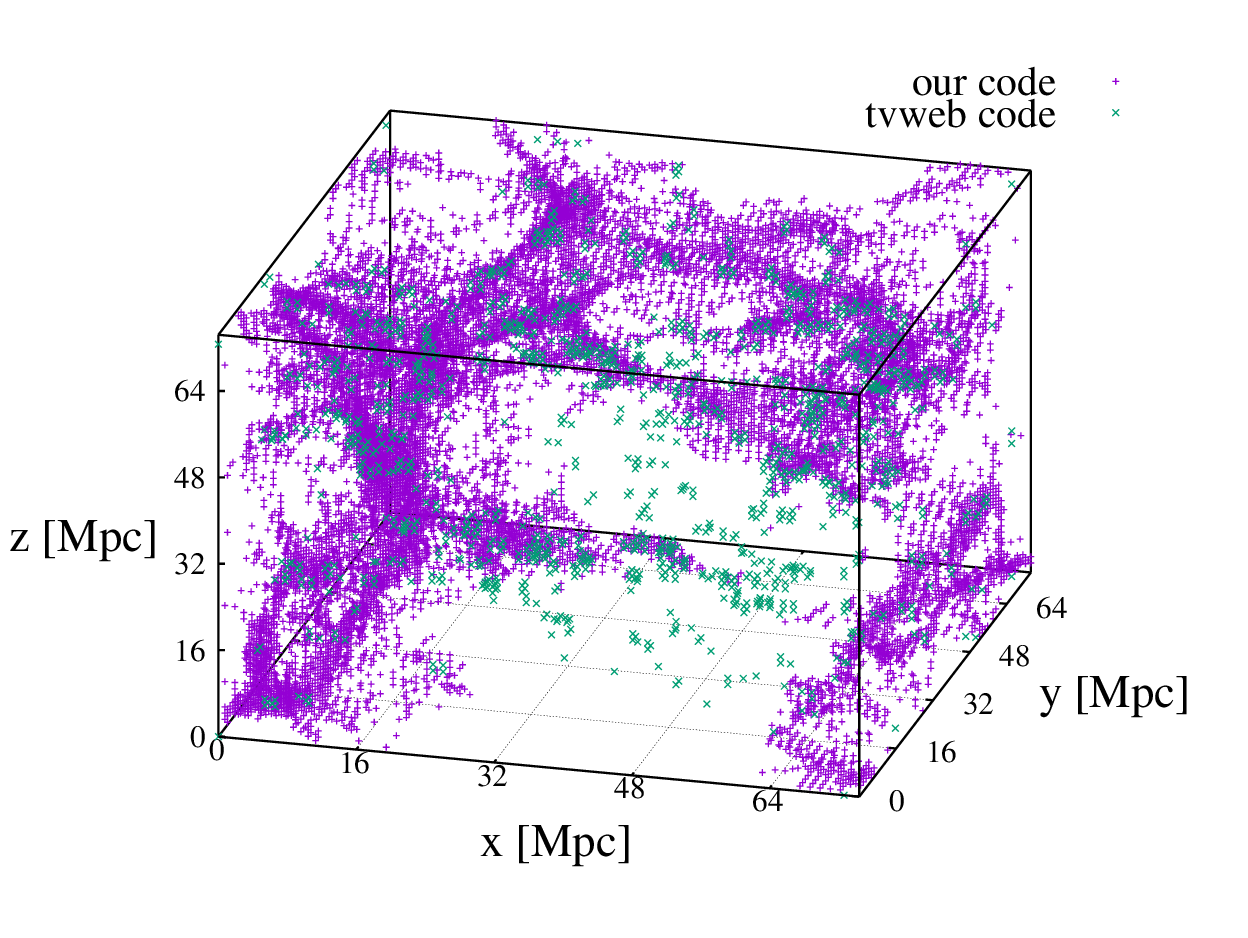}
\caption{\label{CentrosE3} A comparison between the spatial distributions of the cells 
in class 3 according to our code and the TVWEB code.}
\end{center}
\end{figure}

To compare the assigned labels between the TVWEB code and our code, we look at a set of 
15 plots (including samples of each class) and found nine matches and six discrepancies. To further compare 
these methods, we created a 3D plot of all cells in class 3 that were recognises by both codes, as shown in 
Fig.\ref{CentrosE3}. We emphasise that we only plot the centres of the cells, not the simulation particles. Despite the 
difference in the numbers shown above, it can be seen that the spatial distributions of class 3 cells from the two codes are very 
similar. We have observed similar results with other classes of matter structures.  

In addition, differences in class assignment have also been noted in other works; see 
for example \citet{forero}, who made visual corrections to the class 
assignments. Another way to make adjustments to the class assignment was to implement 
a threshold eigenvalue; see \citet{hoffman} and \citet{forero}. In this work, we did not use a
threshold eigenvalue in order not to reduce the number of available plots, as we explain below.
     
According to the results shown in Section \ref{sec:report}, the CNN was successfully trained 
and tested, as the test accuracy (in row 12 of Table \ref{tab:numeroimagenes}) of the CNN is 
0.72 (for model L 0.7, for model S 0.81, for model LS 0.70 and for model LT 0.67). The ability of the CNN to make correct predictions 
for the class of an unseen 
set of plots was found to be 69 percent for models L and S; 38 percent for model LS and 56 bpercent for model LT. We calculate 
the sum of the true positives, as can be seen in row 14 of 
Table \ref{tab:numeroimagenes}. Thus, combining the plots with the two cell sizes in a single set of training plots 
results in the CNN making 
more errors in identifying matter structures. Based on the macro-averaged f1-score, we obtained 0.68, 0.69 and 0.68 
for model L for classes 1, 2 and 3, respectively. For model S, the scores were 0.81, 0.79 and 0.79. For model LS, the scores were 
0.71, 0.70 and 0.68. For model LT the scores were 0.72, 0.72 and 0.72.

It must be emphasised that this level of performance is similar to that of \citet{inoue}, who 
achieved a performance of 64 percent (macro-averaged f1-score) with a CNN for the 
classification of cosmic structures based on galaxies from simulations. \citet{aragon2019} reported an 
accuracy of just over 0.9 for both filaments and walls. The precision values for filaments and 
walls were 0.7 and 0.78, respectively. Similarly, the recall values for filaments and 
walls were 0.78 and 0.77, respectively. In this paper, precision values of just over 0.56 (model L), 0.73 (model S), 
0.61 (model LS) and 0.61 (model LT) were obtained for filaments. For walls, the 
accuracies were 0.52 (model L), 0.71 (model S), 0.55 (model LS)  and 0.59 (model LT). For class 3 structures (clusters) we 
achieved precision and recall values of over 0.95 for all models.     

A second problem is the small number of plots considered in this study. It was found that 
the power of CNNs increases on a logarithmic scale with the size of the dataset; see \citet{sun}. Typically, the 
datasets for CNN models contain tens of thousands of training plots. To partially mitigate this 
situation, we increase the number of plots by rotating the original plots around the origin of the 
coordinates of each cubic cell. 

For the LT model, we create plots of the three projections in the XY, ZX and ZY planes. Let us consider the usefulness of these plots 
for the CNN calculations. As far as we know, there is no way to tell the CNN that the three projections refer to the same 3D matter 
structure, and we believe that the CNN interprets these projections as different examples of labelled structures. For the CNN, plots obtained 
by rotating an original plot multiple times and those obtained by projection are equivalent. For the human eye, there is 
an obvious benefit, as the projections of a 3D structure in the three coordinate planes allow us to get a better idea of 
the overall image, while this is not the case for the CNN.

Therefore, in this paper we managed to run the models with only 
a few hundred plots. However, when we combined the model L and model S, the number of plots 
was unmanageable for the system: 'ArrayMemoryError: Unable to allocate 11.2 GiB for an array 
with shape (3006, 578, 434, 4)'. For this reason, we have reduced the number of plots from the LS model 
that are used as input for the CNN.  

There are also limitations on the number of epochs required to train the CNN, as the training process makes intensive use of 
the RAM memory, which is quickly exhausted. For example, the LT model could be trained for up to 25 epochs (with 1500 plots). In the literature, it 
is common for models to be trained for 60 or more epochs. Although the number of plots varies considerably between models, the results 
are at the same level and indeed similar.

A third problem is that some matter structures with excessive density  
may lie partially outside the cubic cells and the visualisation volume. For this reason, some plots 
have cuts in the overdensity structure. This could be avoided if the overdensity structure is 
located first and then a cubic cell is placed at the corresponding location. In this case, the cover 
mesh would no longer be uniform and other techniques would have to be used.

A fourth point in this paper is to decide on a particular type of CNN with a particular set of 
parameters. We have pointed out in the Subsection \ref{subs:network} that this 
CNN model comes from \citet{bagnato}, who used it to classify 25000 images of dogs and cats (12500 of each class) 
in different sizes. With the parameters given in Table \ref{tab:cnnmodel}, they achieved an accuracy 
of 0.74 in recognising images of dogs and cats. We used an alternative CNN obtained from \citet{bagnato} who used it   
to classify sports in 70000 images showing nine different types of sports. The images had a size of 21x28 
and three colour channels. \citet{bagnato} achieved an accuracy of 0.84 in recognising sports images. Using the same 
parameters as in \citet{bagnato}, we achieved a similar level of performance as the network described in 
Table \ref{tab:cnnmodel}; see appendix \ref{app:network}.  

We used these CNNs to classify images of matter structures (with excessive density) with the same parameters that 
the authors considered. The images they used are very different from our images. Our images also have three color 
channels and they have a width of 578 px and a height of 434 px for model L ( 640 px by 480 px for model S). We did not 
investigated any changes to these parameters. With the current parameter values, an accuracy of about 0.7 was achieved. 

It must be mentioned that the original plots have a bar with a color scale associated with the density values. The blue color 
stands for low-density regions. The green color with medium density regions. The red color stands for high-density regions. The bar 
has been was removed.

\section{Concluding remarks}
\label{sec:conclu}

The formation of matter haloes occurs through the process of gravitational collapse, see 
\citet{arreaga2007,arreaga2016, arreaga2017}. Therefore, 
the plots of haloes must contain information about how this formation process took place. A very homogeneous 
collapse leads to compact, spherical haloes, while a very inhomogeneous collapse
leads to elongated haloes. Between these collapse extremes there is a whole range of possible halo 
configurations. Until a few years ago, such plots could only be analysed by visual inspection.

In this paper, we have looked at the types of plots that are commonly used in particle-based numerical simulations to 
visualise the results. These are mainly two-dimensional Cartesian plots, in particular coloured isodensity plots for a 
slice of particles from the simulation volume. Then, we proposed the goal of this paper, which is to create hundreds of 
plots for a subset of cells of the simulation volume 
to characterise the type of matter structure that each cell contains. Due to the three-dimensional 
nature of the cosmic web, it is almost impossible for the human eye to distinguish the other classes of matter using 
only a two-dimensional Cartesian projection.

Motivated by the success of neural networks in discreminating between images of dogs and cats, the plots mentioned above were used in this 
paper as input for a typical CNN. In Section \ref{subs:network} we have defined several CNN models with layers that have been 
successfully used in other domains to classify images. In the Appendix \ref{app:network}, we look at an alternative CNN. For the snapshot 
considered here at a zero redshift, the axes of the plots vary from -1.5 to 1.5 Mpc for model L (that is, for the visualisation of cell type L) 
and -0.85 to 0.85 Mpc for model S (visualisation of cell type S). For the LS model, we combine the plots of the two previous models. All these 
models use plots with the projection of the XY plane. The LT model uses Cartesian plots of the ZX and ZY planes in addition to the 
projection of the XY plane.

In Section \ref{sec:report} we trained the CNN system for as many epochs as possible, using as many plots as possible. Test accuracy 
values of 60-70 percent were achieved for all models, as can be seen in Table \ref{tab:numeroimagenes}. This is 
a performance level comparable to other models in the literature. The predictive ability 
of this CNN was tested using a confusion matrix, and we found a performance of about 0.69. In general, we found 
that the recognition of class 1 and class 2 structures (filaments and walls) was more problematic and that 
class 3 structures (clusters, also called knots) were more easily recognised. This is to be expected, since the first two types of structure are 
spatially extended, while the latter are spatially compacted. More specific conclusions can be drawn as follows:

\begin{enumerate}

\item It appears that the smallest cell size (type S) provides a better performance in terms of filaments and walls identification.

\item The LS model, which combines plots of different sizes (S and L type cells) gives the worst results. 

\item Although the number of plots varies considerably between the models, the results are at the same level and even similar.

\item All models were able to classify class3 matter structures, which we refer to as clusters. However, when using the f1-score metric, the 
LS and LT models showed a lower level of predictability for class 3 matter structures, as can be seen in Section \ref{sec:reportpredi}.

\item The two CNNs considered in this paper provide generally similar results. However, the CNN in Appendix \ref{app:network} provided 
better results in terms of its ability to predict the class of unseen plots, as shown in row 14 of Table \ref{tab:cnnmodelalter}. Due to 
the smaller number of layers in the alternative CNN, the execution time could be reduced to about half of the execution time of the 
CNN described in Section \ref{subs:network}.

\end{enumerate}

We would like to emphasize that we have not presented this work as an improvement of the methods for 
detecting cosmic web structures, summarized in Section \ref{sec:int}. Nor to present a new result 
of physical interest. Our aim was to show that the plots commonly used to illustrate the results 
of a cosmological simulation contain valuable information about the cosmic structure that can be processed 
using CNN technology.

\citet{siyu} has proposed a model to use the enormous potential of neural networks to 
distinguish between different 
cosmological models. This model is based on linking the Zeldovich initial conditions with an approximation 
to represent the structure evolution in the Universe, which is less costly than using N-body simulations. The 
input data of the neural network consists of 10,000 cubes with the Zeldovich function applied to the 
particles of the simulation. The output data is the matter structure generated by the FastPM tool. The error 
function, to be minimized by the free parameters of the neural network, is the difference 
between the Zeldovich displacement functions for the particles of the simulation. Here it is worth investigating 
the possibility of using a slicing technique and generating plots of the simulation cubes, as we have done in this article, and seeing if 
a label can be assigned to each plot to distinguish between different cosmological models. See also the paper of \citet{hong}.
\appendix

\section{Comparison of the distribution of classes by calculating the Hessian value using two cells of different sizes}
\label{app:compacells}

In Figure \ref{VisCelda} we compare the sizes of the cells considered in this study, i.e. the H-cell, which is used 
to calculate the Hessian, and the L-cell, which is used to visualise the matter structure. In this section, we investigate whether 
there is a significant difference in the set of class labels when we use the L-cell 
instead of the H-cell to diagonalise the Hessian. The mesh for the simulation box has the same centre for both cell types, although 
for the L-cell mesh, many more particles need to be taken into account to compute the Hessian for each cell centre. 

The use of more particles in the Hessian calculation has two consequences: first, the effects of matter outside 
the H-cell are taken into account, which means that the limitation mentioned in Section \ref{subs:cells} when calculating the Hessian 
with the H-cell is reduced; and second, the computation time needed to traverse the simulation box increases significantly.

In Table \ref{tab:comparacioncells}, we present the results of this comparison. In column 1, in rows 1 to 3, nClass $i$ denotes 
the number of cells labelled as class $i$. In the fourth row, $ncells$ stands for the total number of selected cells in the partition. In 
lines 5 to 7, $fclass \, i$ is the proportion of cells of class $i$ to the total number of selected cells in the simulation box. In lines 
8 to 10, $aveC \, i$ is the average number of particles per cell of class $i$. In line 11, the execution time is the time required for a 
serial code to run through all cells of the mesh. In columns 3 and 4 of the Table \ref{tab:comparacioncells}, the value 
in brackets (next to the cell name) indicates the threshold number of particles, i.e. the minimum number of particles that a 
cell must have in order to be selected. 

The most important result of this comparison is that the proportions of cells for each class are very similar, as can be seen in rows 
5, 6 and 7 of columns 3, 4 and 5 of Table \ref{tab:comparacioncells}. Consequently, we expect that the distribution of labels 
will be practically the same whether the H-cell or the L-cell is used to diagonalize the Hessian. It is therefore expected that 
the results of the CNN will be very similar to those already described in Section \ref{sec:results} of this study.

However, in the case of the L-cell, the number of cells per class increases, which means that more plots could be generated for each 
class, as shown in rows 1, 2 and 3 of columns 4 and 5 of Table \ref{tab:comparacioncells}. It is generally known that the larger the 
sample of training plots, the better the results of the CNN. Exploring this possibility will be of interest in future work.

\begin{table}[ph]
\caption{Comparison between cells to calculate the Hessian.}
{\begin{tabular}{|c|c|c|c|c|} \hline
 0 & matter structure &   cell H (150)  & Cell L (150) &  Cell L (300)    \\
1 & nClass 1             &   133            &     817         &        636            \\
2 & nClass 2            &   145            &      998          &        805           \\
3 & nClass 3           & 14515           &     93482         &     75571              \\
4 & ncells               & 14793           &      95297           &      77012             \\
5 & fClass 1           & 0.008991        &  0.008573        &    0.008258                \\
6 & fClass 2           & 0.009802        &   0.010473       &    0.010453               \\
7 & fClass 3           & 0.981207        &    0.980954     &    0.981289               \\
8 & aveC1             &  338              &      1436        &        1781            \\
9 & aveC2             &  373              &      2107        &       2559             \\
10 & aveC3            &  903             &      2525       &       3071              \\
11 & execution time  & 20 Hr           &      60           &       40             \\            
\hline
\end{tabular} }
\label{tab:comparacioncells}
\end{table}

\section{Alternative CNN}
\label{app:network}

\begin{figure}
\begin{center}
\includegraphics[width=4.7 in]{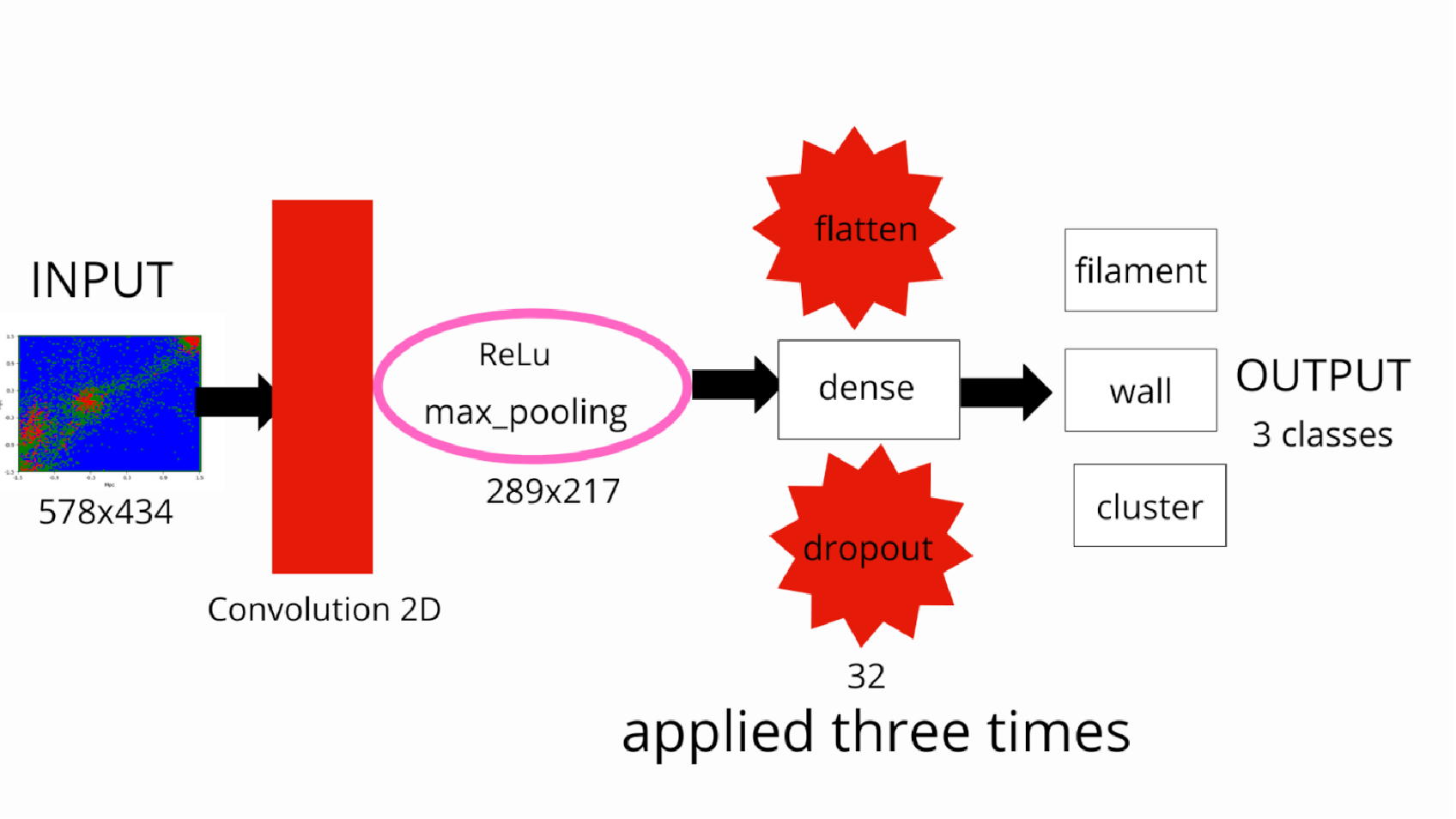} 
\caption{\label{OtraConvolution} Schematic diagram of an alternative CNN. The most important difference between the CNNs is that, in the 
original CNN shown in Fig. \ref{Convolution}, the convolution is applied three times and the output process only once; in the 
alternative CNN of this section, the convolution is applied only once and the output process three times.}
\end{center}
\end{figure}

We have mentioned in Section \ref{sec:dis} that we used an alternative CNN originating from \citet{bagnato}, who used it 
to classify sports in 70000 images with nine different sports. The input plots of this alternative CNN were the same that 
we used in this paper as input for the CNN defined in Section \ref{subs:network}, but the number of plots is slightly 
different due to technical details, which does not change the results significantly. In Fig. \ref{OtraConvolution} we show 
a schematic representation of the architecture of the alternative CNN, which is similar to the one shown in Fig. \ref{Convolution}. 

In fact, the results of these 
two CNNs are very similar, as can be seen in Table \ref{tab:numeroimagenesalter}. It should be noted that 
rows 10-13 of Table \ref{tab:numeroimagenesalter} show the results generated during the code execution where the 
validation data was used. Line 14 shows the results generated with the prediction data. 

In Figure \ref{OtraConvolution}, we show a schematic representation of the architecture of the alternative CNN, for comparison 
with that shown in Figure \ref{Convolution}.
 
\begin{table}[ph]
\caption{Alternative CNN.}
{\begin{tabular}{|c|c|c|} \hline
Layer (type)                   & Output Shape              & Number of Parameters  \\
 conv2d 1  & (None, 578, 434, 32)      & 1184 \\     
 leaky relu 2 & (None, 578, 434, 32)      & 0    \\     
 max pooling 2d 1  & (None, 289, 217, 32)     & 0    \\     
 dropout 2         & (None, 289, 217, 32)      & 0    \\    
 flatten 1          & (None, 2006816)           & 0     \\    
 dense 2              & (None, 32)                & 64218144 \\
 leaky relu 3    & (None, 32)                & 0  \\       
 dropout 3         & (None, 32)                & 0   \\     
 dense 3             & (None, 3)                 & 99  \\     
\hline
\end{tabular} }
\label{tab:cnnmodelalter}
\end{table}                   

\begin{table}[ph]
\caption{The number of images used and some results.}
{\begin{tabular}{|c|c|c|c|c|c|c|} \hline
row &                   & Model L         &  Model S       &  Model LS    &   Model LT \\
1   &Class 1          &   462           &    762         &   610                &     600       \\
2   &Class 2          &   466           &    834         &   662                &     600          \\
3   &Class 3          &   433           &    801         &   624                &     300         \\
4   &training plots   &   1088          &   1917         &  1516            &     1200        \\
5   &validation plots &   273           &    480         &   304             &     300          \\
6   &prediction plots &    90           &     90         &   180              &      180          \\
7   &total plots      &  1451           &   2397         &  1896             &       1500        \\
8   &Total parameters &  64219427       &  78644515      &  64219427     &   78644515        \\
9   &Memory           &  244.98 MB      &   300 MB       &  244.98 MB       &    300 MB      \\
10  & correct labels   &  186            &   323          &   267                     &     193           \\
11  &incorrect labels  &   87            &   157          &   113                     &      107  \\
12  &  test accuracy   &   0.68          &  0.67          &   0.70                   &      0.64           \\
13  &   test loss      &   0.76          &  0.67          &   0.59                      &       0.58           \\
14  & true positives   & 0.22,0.18,0.31  & 0.26,0.13, 0.33&  0.21, 0.13, 0.30   &  0.11,0.21,0.13   \\
15  & $\Delta_L$       &   1.5 Mpc  &  0.85 Mpc &  1.5 and 0.85 Mpc  &   1.5 Mpc       \\
\hline
\end{tabular} }
\label{tab:numeroimagenesalter}
\end{table}

\begin{acknowledgements}
The author gratefully acknowledges the computer resources, technical expertise, and
support provided by the Laboratorio Nacional de Superc\'omputo del Sureste de M\'exico
through grant number O-2016/047. We also appreciate the collaboration with the 
physics students Jos\'e Antonio Sanabria V\'azquez and Carlos Oswaldo Ochoa Boj\'orquez.
\end{acknowledgements}

\end{document}